\documentclass[journal,10pt,twocolumn,journal,english,american]{IEEEtran}
\usepackage{url,graphicx,myPack,myMathPack,algorithm2e}

\usepackage{color}
\definecolor{FireBrick}{rgb}{0.5812,0.0074,0.0083}
\definecolor{RoyalBlue}{rgb}{0.0236,0.0894,0.6179}
\definecolor{RoyalGreen}{rgb}{0.0236,0.6179,0.0894}
\definecolor{RoyalRed}{rgb}{0.6179,0.0236,0.0894}
\definecolor{LightBlue}{rgb}{0.8544,0.9511,1.0000}
\definecolor{Black}{rgb}{0.0,0.0,0.0}

\definecolor{linkColor}{rgb}{0.0,0.0,0.554}
\definecolor{citeColor}{rgb}{0.0,0.0,0.554}
\definecolor{fileColor}{rgb}{0.0,0.0,0.554}
\definecolor{urlColor}{rgb}{0.0,0.0,0.554}
\definecolor{promptColor}{rgb}{0.0,0.0,0.589}
\definecolor{brkpromptColor}{rgb}{0.589,0.0,0.0}
\definecolor{gapinputColor}{rgb}{0.589,0.0,0.0}
\definecolor{gapoutputColor}{rgb}{0.0,0.0,0.0}

\definecolor{FuncColor}{rgb}{0.0,0.0,0.0}
\definecolor{Chapter }{rgb}{0.0,0.0,0.0}
\definecolor{DarkOlive}{rgb}{0.1047,0.2412,0.0064}
\usepackage{fancyvrb}

\begin{document}

\title{Explicit Polyhedral Bounds on Network Coding Rate Regions via Entropy Function Region: Algorithms, Symmetry, and Computation}
\author{Jayant Apte,~\IEEEmembership{Student Member, IEEE,} and %
John MacLaren Walsh,~\IEEEmembership{Senior~Member, IEEE.}
\thanks{%
Support under National Science Foundation awards CCF--1016588 and 1421828 is gratefully acknowledged.
}%
\thanks{%
J. Apte and J.~M.~Walsh are with the Department of Electrical and Computer Engineering, Drexel University, Philadelphia, PA USA (email: \textsf{jsa46@drexel.edu}, and \textsf{jwalsh@coe.drexel.edu}).
Preliminary results were presented at ISIT 2015 \cite{JayantISIT2015} and NetCod 2015 \cite{JayantNetCod2015}.
}%
}
\maketitle
\noindent\begin{abstract}
Automating the solutions of multiple network information theory problems, stretching from fundamental concerns such as determining all information inequalities and the limitations of linear codes, to applied ones such as designing coded networks, distributed storage systems, and caching systems, can be posed as polyhedral projections.  These problems are demonstrated to exhibit multiple types of polyhedral symmetries.  It is shown how these symmetries can be exploited to reduce the complexity of solving these problems through polyhedral projection.
\end{abstract}

\section{Introduction}

Converse proofs in network information theory consist of sequences of inequalities derived by applying fundamental laws, such as the positivity of conditional entropy, mutual information, and conditional mutual information, as well as problem constraints, including any specified independences and Markov chains created from encoding and decoding constraints \cite{YeungBook,YoungHanKimBook}.  The aim of these sequences of inequalities is to obtain regions that interrelate exclusively those quantities of interest, typically including various encoding rates and fidelity criteria for lossy or lossless reproduction.  While intermediate rate region expressions presented in theorems often involve auxiliary random variables, not explicitly included in the problem, whose distributions are left free other than to obey a certain series of constraints, the region of interest is only ultimately obtained after optimization over the distribution of these auxiliaries \cite{Blahut_TIT_4_72}, yielding an ultimate region relating exclusively the quantities of interest.

It remains relatively uncommon to recognize proving a converse or outer bound in network information theory as a polyhedral projection, yet this process is indeed mathematically equivalent to polyhedral projection.  This fact has been made clearest in the context of network coding (broadly defined) problems \cite{YeungBook}, including determining the capacity regions for multi-source network coding \cite{yanmultisourceTIT}, as well as storage repair tradeoffs for distributed storage systems with exact repair \cite{TianJSAC433}, and cache size vs. rate during the delivery phase in coded caching \cite{ChaoCache}.  From a more fundamental standpoint, proving information inequalities \cite{Zhang_TIT_07_98,BookInequalities}, as well as those inequalities that linear codes must obey \cite{Kinser2011NewIneqSubspaceArra,DFZ2009Ineqfor5var}, have also been posed in a manner reminiscent of polyhedral projection.  \S \ref{sec:convProofExamp} of this paper provides a series of examples demonstrating of each of these problems can be explicitly formulated as polyhedral projection problems.  

A key benefit of formulating rate region calculation problems as polyhedral projection problems is that it enables them to be solved via computer algorithms rather than undergoing the arduous process of deriving them by hand.  In \S \ref{sec:convProofExamp} we draw a significant distinction between \emph{verification}, which, given a putative inequality among the quantities of interest, provides a sequence of inequalities deriving it, and that of exhaustive \emph{generation} which determines a minimal set of all inequalities relating exclusively the quantities of interest.  We review that verification can be posed as a linear program, while generation is posed as a polyhedral projection.  Prior computational toolkits for information theory such as the information theoretic inequality prover (ITIP) \cite{itip}, xITIP \cite{xitip}, and miniITIP \cite{minitip} have aimed at \emph{verification}, while the toolkit accompanying this manuscript, the \emph{information theoretic converse prover} ITCP performs exhaustive generation by implementing the highly efficient form of polyhedral projection described within.

From a computational perspective, proving converses in network information theory via polyhedral projection is only feasible for relatively small problems, as the common formulation involves a projecting a polyhedron with a number of indeterminates that is exponential in the number of sources and messages.  This implies that in order to push computer derived proofs in network information theory as far and to as large problems as possible, the polyhedral projection algorithm should be designed to reduce the complexity of the projection process as much as possible.  An important first step in this direction, noted by other authors \cite{Xu_ISIT_08,CsirmazBenson} in the particular context of deriving non-Shannon information inequalities, is to utilize methods such as the Convex Hull Method \cite{lassezchm,jayantchm} or Benson's algorithm \cite{CsirmazBenson,benson1998}, which work directly in the projection space by solving carefully designed linear programs over the polyhedron to project, rather than Fourier Motzkin \cite{PermuterFME,HuynhLL92} which suffers from substantially higher complexity owing to its incremental removal of dimensions.  

This manuscript focuses on a second key aspect of reducing the complexity of proving network information theory converses via polyhedral projection: exploiting symmetry.  First, \S \ref{sec:typesOfSymmetry} reviews several different notions of symmetry for polyhedra, then describes their application both the polyhedra to project, and the resulting answer, for proving each of the types of information theoretic outer bounds given as examples in \S \ref{sec:convProofExamp}.  Next, \S \ref{sec:symCHM} describes three distinct ways known symmetry groups can be exploited in polyhedral projection, forming a new algorithm for symmetry exploiting polyhedral projection symCHM.  The complexity reductions of polyhedral projection afforded by each of these three symmetry techniques is illustrated and quantified in the context of specific example network information theory problems from \S \ref{sec:convProofExamp}.  Finally, \S \ref{sec:ITCP} describes a software package for the GAP system \cite{GAP4}, ITCP \cite{ITCPsoftware}, which accompanies the manuscript and implements the methods described in \S \ref{sec:symCHM}, providing examples showing how to utilize it to calculate outer bounds for network information theory problems.  A parallel series of results and algorithms in a companion manuscript \cite{Apte_NCRR}, and a second software package, the \emph{information theoretic achievability prover} (ITAP) \cite{ITAPsoftware} take up the issue of providing achievability proofs to match the converse proofs derived via techniques in this manuscript.  Together, ITCP and ITAP have been used to determine the rate regions of millions of new network information theory problems \cite{Congduan_MDCS,Li_Operators} and a new theory relating network coding problems of different sizes has been developed to best harness them to solve network information theory problems of arbitrary scale \cite{Li_Operators}.

\section{Converse Proofs in Network Information Theory are Polyhedral Projections}\label{sec:convProofExamp}
This section identifies proving converses for a series of network information theory problems as polyhedral projection.
\subsection{Network Coding Rate Regions}\label{sec:netCod}
\begin{figure}
\centerline{\includegraphics[scale=0.4]{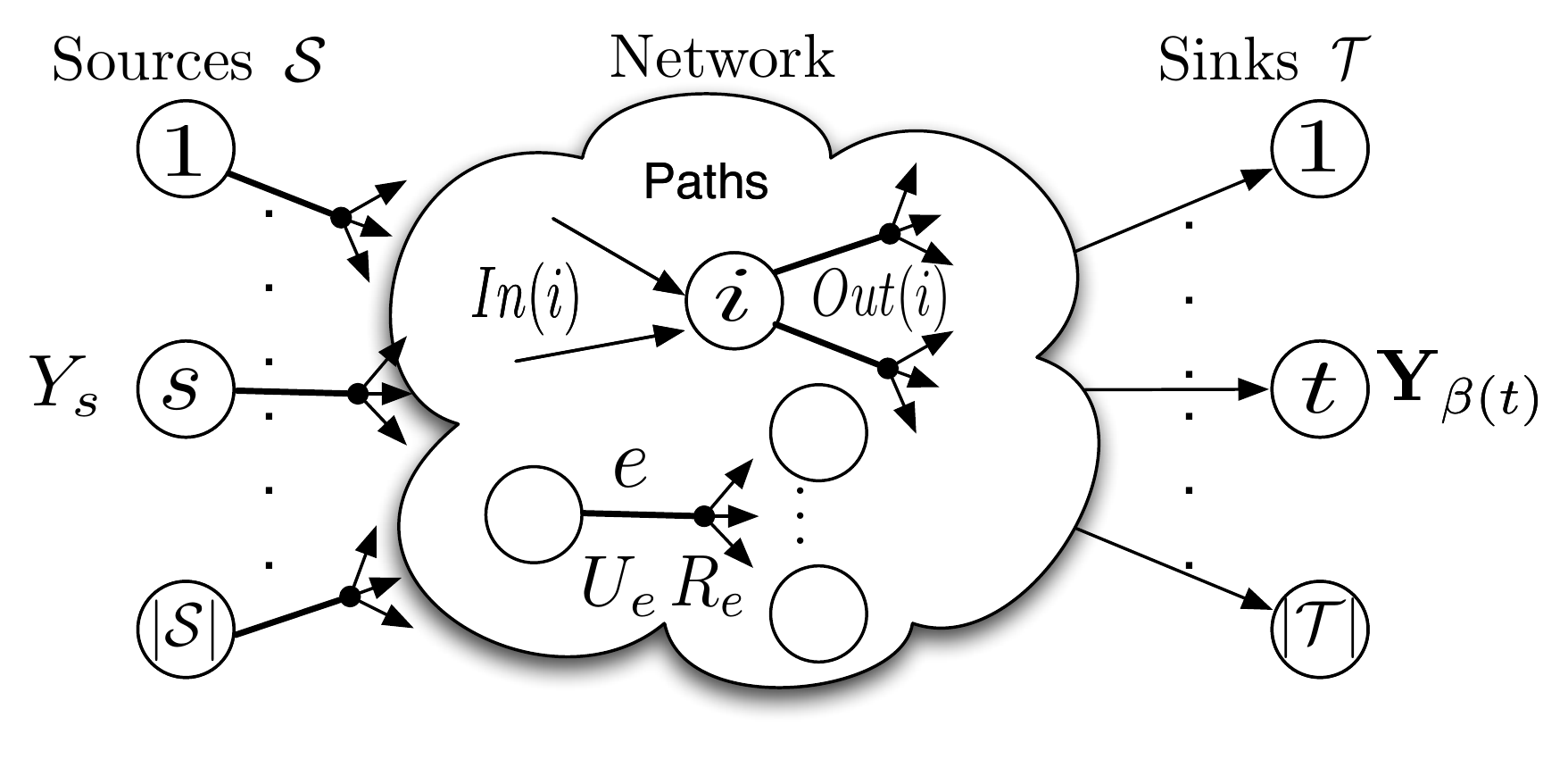}}
\caption{\label{fig:generalnetwork}A general network model $\textsf{A}$.}
\vspace{-0.4cm}
\end{figure}
Following \cite{Li_Operators} which builds upon \cite{yanmultisourceTIT} and \cite{YeungBook}, a network coding problem consists of a labeled directed acyclic hypergraph $(\mathcal{V},\mathcal{E})$ as in Fig.\,\ref{fig:generalnetwork}, consisting of a set of nodes $\mathcal{V}$ and a set $\mathcal{E}$ of directed hyperedges in the form of ordered pairs $e=(v,\mathcal{A})$ with $v\in \mathcal{V}$ and $\mathcal{A} \subseteq \mathcal{V} \setminus v$.  The nodes can be partition into three sets $\mathcal{V}=\mathcal{S} \cup \mathcal{G} \cup \mathcal{T}$: the sources $\mathcal{S}$ which have no incoming edges and exactly one outgoing edge, the intermediate nodes $\mathcal{G}$ which have incoming and outgoing edges, and the sinks $\mathcal{T}$ which no outgoing edges.  The outgoing edge from each source node $s\in \mathcal{S}$ carries a source random variable $Y_s$ with entropy $H(Y_s)$.  Each outgoing edge $e\in \mathcal{E},\ e=(v,\mathcal{A})$ from an intermediate node $v \in \mathcal{G}$ carries a message random variable $U_e$ that must be encoded exclusively from the messages on those edges $\textrm{In}(e)=\textrm{In}(v) = \left\{ e' \in \mathcal{E} | e'=(v',\mathcal{A}'),\ v \in \mathcal{A}'\right\}$ coming into $v$.  The sinks are labeled with a demand function $\beta:\mathcal{T} \rightarrow 2^{\mathcal{S}}$ with $\beta(t), t\in \mathcal{T}$ indicating the subset of sources which must be decodable from the messages on edges $\textrm{In}(v)$ incoming at $t$.  While this is the most common and straightforward representation of a network coding problem, it enables a substantial redundancy and un-necessary components to be included, and hence \cite{Li_Operators} presents both a more concise problem representation and a series of minimality conditions removing un-necessary components that are extraneous to the model.

Defining $\mathcal{L}_{i},i=1,3,4',5$ as network constraints representing source
independence, coding by intermediate nodes, edge capacity constraints, and sink nodes decoding constraints respectively,
\begin{IEEEeqnarray*}{rCl}
\mathcal{L}_{1} & = & \{[\mathbf{h}^T,\boldsymbol{r}^T]^T :h_{\boldsymbol{Y}_{\mathcal{S}}}=\Sigma_{s\in\mathcal{S}}h_{Y_{s}}\}  \label{eq:rrcondef1} \\
\mathcal{L}_{3} & = & \{[\mathbf{h}^T,\boldsymbol{r}^T]^T:h_{\boldsymbol{U}_{{\rm Out}(g)}|(\boldsymbol{Y}_{\mathcal{S}\cap\mathrm{In}(g)}\cup \boldsymbol{U}_{\mathcal{E}_U\cap\mathrm{In}(g)})} =0,g\in \mathcal{G}\} \label{eq:rrcondef2} \\
\mathcal{L}_{4'}&=&\{[\mathbf{h}^T,\boldsymbol{r}^T]^T :R_e\geq h_{U_{e}}, \forall e\in\mathcal{E}_U\} \label{eq:Lratefree}\\
\mathcal{L}_{5} & = & \{[\mathbf{h}^T,\boldsymbol{r}^T]^T :h_{\boldsymbol{Y}_{\beta(t)}|\boldsymbol{U}_{\text{In}(t)}}=0,\forall t\in\mathcal{T}\} \label{eq:rrcondef4},
\end{IEEEeqnarray*}
and denoting $\mathcal{L}_{13} = \mathcal{L}_1 \cap \mathcal{L}_3$, $\mathcal{L}_{4'5} = \mathcal{L}_{4'}\cap\mathcal{L}_5$ and $\mathcal{L}_{\mathsf{A}} = \mathcal{L}_1  \cap \mathcal{L}_3 \cap \mathcal{L}_{4'} \cap \mathcal{L}_5$, a seminal result of Yan, Yeung, and Zhang \cite{YeungBook,yanmultisourceTIT} can be translated \cite{Li_PhDdissertation,CongduanTranIT2015Arxiv} to the following theorem.

\begin{theorem}[\cite{YeungBook,yanmultisourceTIT}]
\label{thm:rateregion}
The rate region of a network $\mathsf{A}$ is expressible as
\begin{equation}
\mathcal{R}_{c}(\mathsf{A})=\mathrm{Proj}_{\boldsymbol{r},\boldsymbol{\omega}}(\overline{\rm{con}(\Gamma_{N}^{*}\cap\mathcal{L}_{13})}\cap\mathcal{L}_{4'5}),\label{eq:generalrateregionfree}
\end{equation}
where ${\rm con}(\mathcal{B})$ is the conic
hull of $\mathcal{B}$, and $\mathrm{Proj}_{\boldsymbol{r},\boldsymbol{\omega}}(\mathcal{B})$
is the projection of the set $\mathcal{B}$ on the coordinates $\left[\boldsymbol{r}^T,\boldsymbol{\omega}^T \right]^T$ where $\boldsymbol{r} = \left[ R_e | e\in\mathcal{E}_U\right]$ and $\boldsymbol{\omega} = \left[H(Y_s) | s\in\mathcal{S}\right]$.
\end{theorem}

This is only an implicit characterization as it involves the region of entropic vectors $\Gamma^*_N$ \cite{YeungBook}, which is unknown and even non-polyhedral \cite{Matus_ISIT_2007} for $N\geq 4$.  However, a key motivation behind the result is that $\Gamma^*_N$ can be replaced with a polyhedral outer bound $\Gamma^{\textrm{out}}_N$, e.g. the Shannon outer bound \cite{YeungBook}
\begin{equation*}
\Gamma_N := \left\{ \boldsymbol{h} \in \mathbb{R}^{2^{\mathcal{N}}\setminus \emptyset} \left| \begin{array}{c} h_{\mathcal{N}} - h_{\mathcal{N}\setminus \{i\}} \geq 0 \\
h_{i\mathcal{K}} + h_{j\mathcal{K}} - h_{\mathcal{K}} - h_{ij\mathcal{K}} 
\geq 0 \\
\forall i,j \in \mathcal{N}\ \forall \mathcal{K} \subseteq \mathcal{N} \setminus \{i,j\} \end{array}  \right. \right\},
\end{equation*}
obtaining the outer-bound $\mathcal{R}_o \supseteq \mathcal{R}_c$ via the polyhedral projection
\begin{equation}\label{eq:netCodRateRegion}
\mathcal{R}_o = \mathrm{Proj}_{\boldsymbol{r},\boldsymbol{\omega}} \left(\Gamma^{\textrm{out}}_N \cap\mathcal{L}_{\mathsf{A}} \right)
\end{equation}
It is precisely in this manner that a generating a converse proof in network coding can be viewed as a process of polyhedral projection.  An \emph{explicit polyhedral outer bound} (EPOB) to the network coding rate region is a description of $\mathcal{R}_o$ as a series of inequalities or rays involving exclusively the rate dimensions $\mathbf{r}$ and $\mathbf{\omega}$.  The resulting converse proofs track which inequalities and equalities in the parent polyhedron $\Gamma^{\textrm{out}}_N \cap\mathcal{L}_{\mathsf{A}}$ which must be summed to obtain each of the inequalities describing the projection $\mathcal{R}_o$.

\begin{figure}
\begin{center}
\includegraphics[scale=.7]{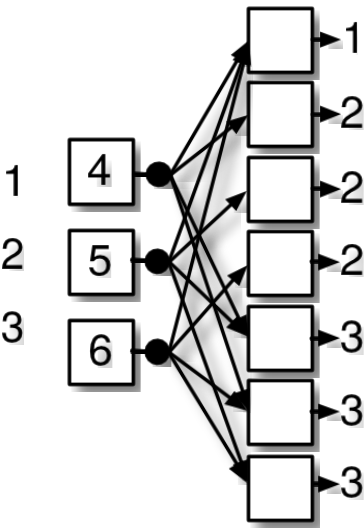}
\end{center}
\caption{The IDSC instance $\textsf{A}$, with 3 source messages labeled $1,2,3$ and 3 encoders whose output messages are labeled labeled $4,5,6$. On the extreme right hand side are $7$ decoders, each of which demands a subset of the source messages.}\label{fig:idsc_ex}
\end{figure}
The next example solidifies these concepts.
\exmpl{[Network Coding Rate Region]  Consider the network coding problem depicted in Fig. \ref{fig:idsc_ex}, in which a series of three sources $1,2,3$ are all made to available to each of three encoders $4,5,6$, each of which creates single message which is then overhead by a subset of decoders, with each decoder demanding only one of the three sources.  This type of network coding problem with no intermediate nodes is called an independent distributed source coding (IDSC) problem.  The set of constraints $\mathcal{L}_{\mathsf{A}}$ for this problem are
\begin{equation} \label{eq:ncCons}
\begin{aligned}
h_1+ h_2+ h_3 &=  h_{1,2,3} ,&  h_{1,2,3,4,5,6} &=  h_{1,2,3}\\
 h_{1,4,5,6} & =  h_{4,5,6} ,&  h_{2,4} &=  h_4\\
 h_{2,5} &=  h_5 ,&  h_{2,6} &=  h_6\\
 h_{3,4,5} &=  h_{4,5} ,&  h_{3,4,6} &=  h_{4,6}\\
&  h_{3,5,6} =  h_{5,6} & 
\end{aligned}
\end{equation}
With $\Gamma_{\textsf{out}}$ defined as the intersection of all Shannon-type information inequalities for $6$ discrete random variables, the associated EPOB $\mathcal{R}_o$ is given by the following inequalities
\begin{equation}\label{eq:epobex}
\begin{aligned}
\omega_1 &\geq 0 \\
\omega_2 &\geq 0  \\
\omega_3 &\geq 0  \\
\omega_2 &\leq R_4\\
\omega_2 &\leq R_5\\
\omega_2 &\leq R_6\\
2\omega_2+\omega_3 &\leq R_4+R_5\\
2\omega_2+\omega_3 &\leq R_4+R_6\\
2\omega_2+\omega_3 &\leq R_5+R_6\\
2\omega_1+6\omega_2+3\omega_3 &\leq R_4+R_5+R_6\\
\omega_1+4\omega_2+2\omega_3 &\leq 2R_4 + R_5 + R_6\\
\omega_1+4\omega_2+2\omega_3 &\leq R_4 + 2R_5 + R_6\\
\omega_1+4\omega_2+2\omega_3 &\leq R_4 + R_5 + 2R_6\\
\end{aligned}
\end{equation}
}{ex:run}
We will use this problem as a running example throughout the manuscript to illustrate the concepts, and we will demonstrate the algorithms we will describe can generate the rate region description EPOB (\ref{eq:epobex}) from the problem description shown in Fig. \ref{fig:idsc_ex}.

\subsection{Storage Repair Tradeoff in Exact Repair Distributed Storage Systems}
Of substantial recent research interest has been the tradeoff between the amount that can be stored and the amount of information that must be transferred to repair a failed disk in certain coded designs for very large distributed information storage systems \cite{dimakis10}.  By drawing graphs with receivers representing different decoding conditions and different repair conditions the storage repair tradeoff can be viewed as being derived from the capacity region of a network coding problem, with the twist in exact repair that some decoders must reproduce not original sources, but instead edges in the computations\cite{TianJSAC433}.  Nevertheless, utilizing the constraints $\mathcal{L}_{\mathsf{A}}$ described in the previous section with the adaptation of reproduction of edges, and a common rate limitation for edges of the same type (storage or repair), one can obtain the storage repair tradeoff as an instance of a polyhedral projection problem \cite{TianJSAC433}.  While must of the research on these problems is typically focussed on a single source representing all of the information to be stored and an associated two dimensional tradeoff between storage and repair, multisource formulations \cite{Apte_CISS_2014}, for instance reflecting data with different latency requirements, or heterogenous storage sizes and repair bandwidths, are possible and can be derived as polyhedral projections of largely the same form as (\ref{eq:netCodRateRegion}).

\subsection{Cache Delivery Tradeoff in Coded Caching Systems}
A similar variant of network coding problems draws motivation of the possibility of storing encodings across high demand content at wireless basestations or receivers during off peak hours with the goal of reducing traffic on broadcast links during peak hours.  A natural tradeoff studies the size of the cache and what must be broadcasted under worst case content demands.  By drawing appropriate hypergraphs with different sources representing different content to request, and different hyperedges representing cache contents and the broadcasted common message under different content requests, and with a common rate limitation for the broadcast messages, and a separate common rate limitation for the cache messages  \cite{ChaoCache}, it can be seen that these problems fall into the class of network coding problems described in \S \ref{sec:netCod}.  Determining their tradeoffs can thus be posed as polyhedral projection problems of the form in (\ref{eq:netCodRateRegion}).

\subsection{Non-Shannon Information Inequalities}
The most common way to generate tighter outer bounds on the entropy region $\bar{\Gamma}^*_N$ for $N\geq 4$ is through polyhedral projection of a constrained Shannon outer bound in a larger series of variables \cite{Xu_ISIT_08,BookInequalities,Kaced13}.  Indeed, given a collection of $N$ random variables in which one wants to derive the new non-Shannon inequality, one augments them with $k$ new ``constrained'' random variables generated according to a ``copy'' construction of being independent of some random variables while having a matching the conditional distribution with some others \cite{Xu_ISIT_08,BookInequalities}.  These copy constraints imply linear equalities between certain subset entropies in all of the $N+k$ random variables, which are intersected with $\Gamma_{N+k}$ or a polyhedral entropy outer bound created in a previous step in the process.  Then the entropies containing any of the $k$ constrained random variables are dropped via polyhedral projection.  The resulting cone is still an outer bound to entropy, but can be smaller than the Shannon outer bound.  As such, the process of proving non-Shannon inequalities itself can be posed as a polyhedral projection in which a Shannon outer bound is projected after intersecting it with linear constraints.

\subsection{Linear Rank Inequalities}
Inequalities for the dimensions of Minkowski sums of subsets of a set of subspaces can be derived through a similar process of polyhedral projection \cite{Hammer2000ShannEntr,KinserSubspace,DFZ2009Ineqfor5var}.  These inequalities reflect extra constraints that the rates and entropies of messages created through linear network codes must obey \cite{DFZ2009Ineqfor5var,DFZ_Insuff}.  The most common method utilized to generate new such inequalities observes that random variables that are created through linear transformations (over a finite field) of uniformly distributed random variables (over the same finite field), must have \emph{common information}.  In particular, if $X$ and $Y$ are two random variables associated with such a linear construction, then there exists a random variable $Z$, the \emph{common information} such that $H(Z|X)=H(Z|Y) = 0$ and $H(Z) = I(X;Y)$ \cite{Hammer2000ShannEntr,DFZ2009Ineqfor5var}.  Forcing the existence of common information restricts the type of distributions that are possible for $X$ and $Y$, but these include all $X$ and $Y$ created through such a linear construction.  Thus, one can obtain outer bounds on the region of entropic vectors (and hence subspace dimensions) that be be achieved via such linear transformations from uniforms via the following process.  One again considers $N$ such subspaces, then creates $k$ random variables representing such common informations.  One then intersects series of inequalities that must be obeyed by such rank vectors, which for instance include the Shannon type inequalities $\Gamma_{N+k}$, with the equalities (cases of $H(Z|X)=H(Z|Y) = 0$ and $H(Z) = I(X;Y)$) implied by the definition of the $k$ common informations.  Then, one projects out any dimensions associated with the common information variables.  The resulting polyhedral cones inequalities hold for any $N$ subspace ranks.  This process, together with subspace constructions achieving each extreme ray, has been utilized to determine all of the inequalities among subset-sum ranks of $N\leq 5$ subspaces \cite{Hammer2000ShannEntr,DFZ2009Ineqfor5var}, and is at present being used by several scholars to study the cases $N\geq 6$.  By putting them as outer bounds (not for entropy in general, but those entropies achievable with linear codes) in (\ref{eq:netCodRateRegion}), one can obtain outer bounds on the region of rates achievable with linear network codes.


\section{Polyhedral Symmetries in Network Information Theory} \label{sec:typesOfSymmetry}
Each of the polyhedral bounds to multiterminal information theory problems obtained through polyhedral projection as reviewed in \S \ref{sec:convProofExamp} can exhibit high amounts of symmetry, which in turn may be exploited to reduce the complexity of obtaining an explicit polyhedral outer bound.  In this section, we first present in \S \ref{sec:polyNot} and \S \ref{sec:polySym} formal definitions for various notions of symmetry for a polyhedron, and then demonstrate in \S \ref{sec:entSym} and \S \ref{sec:netCodSym} how these definitions apply to the types of polyhedral cones encountered in the network information theory polyhedral projection problems detailed in \S \ref{sec:convProofExamp}.
%
\subsection{Related Concepts from Polyhedra}\label{sec:polyNot}
\newcommand{\homog}{\textrm{homog}}
\newcommand{\polyhedron}{\mathcal{P}}
\newcommand{\extremePointSet}{\mathcal{S}}
\newcommand{\extremeRaySet}{\mathcal{T}}
\newcommand{\extremePoint}{\mathbf{s}}
\newcommand{\extremeRay}{\mathbf{t}}
\newcommand{\polyCone}{\mathcal{C}}
\newcommand{\combSymGroup}{\mathsf{CSG}}
\newcommand{\coneRays}{\mathcal{T}}
\newcommand{\coneRay}{\mathcal{R}}
\newcommand{\coneRayRep}{\mathbf{v}}
\newcommand{\coneRayReps}{\mathcal{V}}
\newcommand{\rayPerm}{\psi}
\newcommand{\coneRaysSub}{\mathcal{A}}
\newcommand{\affineSymGroup}{\mathsf{ASG}}
\newcommand{\generalLinearGroup}{\mathbb{GL}}
\newcommand{\resSymGroup}{\mathsf{RSG}}
\newcommand{\permSymGroup}{\mathsf{PSG}}
\newcommand{\linearTrans}{\mathbb{A}}
\newcommand{\face}{\mathcal{F}}

Polyhedra have several important cascading notions of symmetry formalized through groups.  To properly define these groups, notation for several important concepts in polyhedra must be fixed.  A polyhedron is typically expressed as the set of solutions to a system of linear inequalities
\begin{equation}
\polyhedron = \left\{ \mathbf{x} \in \mathbb{R}^d | \mathbb{A}' \mathbf{x} \geq \mathbf{b}' \right\}
\end{equation}
for some $\mathbb{A}' \in \mathbb{R}^{M\times d}$ and $\mathbf{b}' \in \mathbb{R}^{M}$.  When working computationally with polyhedra, it is common to consider exclusively polyhedral cones, for which $\mathbf{b}' = \mathbf{0}$, as when $\mathbf{b}'$ is non-zero we can pass to the \emph{homogenization}
\begin{equation}
\homog(\polyhedron) = \left\{ \left[\begin{array}{c} x_0 \\ \mathbf{x} \end{array} \right] \in \mathbb{R}^{d+1} \left| -x_0 \mathbf{b}' + \mathbb{A} \mathbf{x}' \geq \mathbf{0} \right. \right\}
\end{equation}
which is a polyhedral cone of one higher dimension.  

A face $\face$ of a polyhedral cone $\polyCone$ is a convex subset $\face \subset \polyCone$ such that any line segment in $\polyCone$ with a relative interior point in $\face$ must also have both of its end points in $\face$ \cite{Rockafellar_Book}.  A polyhedral cone $\polyCone$ is said to be pointed if it contains no lines, and the discussion for the remainder of the manuscript refers to pointed polyhedral cones.  

A face of a pointed polyhedral cone that is one dimensional is called an \emph{extreme ray} $\coneRay$ of the cone, and can be represented with a vector $\coneRayRep$ as $\coneRay = \left\{\gamma\coneRayRep | \gamma \geq 0\right\}$.  Pointed polyhedral cones have a finite number of such extreme rays, and thus in addition to being represented as the set of solutions to a finite homogeneous system of linear inequalities
\begin{equation}
\polyCone = \left\{ \mathbf{x} \in \mathbb{R}^d \left| \mathbb{A} \mathbf{x} \geq \mathbf{0} \right. \right\}
\end{equation}
with $\mathbb{A}$ having linearly independent columns, pointed polyhedral cones can also be represented as the conic hull of a finite set $\coneRayReps$ containing representatives of the set of their extreme rays $\coneRays$
\begin{equation}
\polyCone =\textrm{conic}(\coneRays) = \left\{ \sum_{\coneRayRep \in \coneRayReps} \gamma_{\coneRayRep} \coneRayRep \left| \gamma_{\coneRayRep} \geq 0\  \forall \coneRayRep \in \coneRayReps \right. \right\}.
\end{equation}
The matrix representatives $\mathbb{A}$ for the inequalities and $\mathbb{V}=[\coneRayRep | \coneRayRep \in \coneRayReps]^T$ for the extreme rays are thus known as a \emph{double descriptions pair}.
Likewise, each higher dimensional face $\face$ of a pointed polyhedral cone $\polyCone$ can be represented as the conic hull $\face = \textrm{conic}(\coneRaysSub)$ of a certain subset $\coneRaysSub \subseteq \coneRayReps$ of the extreme ray representatives.  The set of faces of the cone form a lattice via the containment relation called the face lattice\cite{FukudaPolyCompNotes}.

\subsection{Symmetry Groups of Polyhedra}\label{sec:polySym}
The most expansive and general notion of symmetry for a polyhedral cone $\polyCone$ is the \emph{combinatorial symmetry group}.
\defn{[Combinatorial Symmetry Group] The combinatorial symmetry group $\combSymGroup(\polyCone)$ of a polyhedral cone is the subgroup of the symmetric group of order $| \coneRays |$ consisting of those permutations (bijections $\rayPerm:\coneRays \rightarrow \coneRays$) of the set extreme rays that leave the face lattice invariant.} 
In other words, $\rayPerm$ is a combinatorial symmetry if, for any subset $\coneRaysSub \subseteq \coneRays$ whose conic hull $\textrm{conic}(\coneRaysSub)$ is a face of $\polyCone$, $\rayPerm(\coneRaysSub) = \left\{ \rayPerm(\coneRay) | \coneRay \in \coneRaysSub \right\}$ is also a face of $\polyCone$.

While the combinatorial symmetry group is the most general notion of symmetry of a polyhedral cone, since it is represented as a subgroup of the symmetric group of order $|\coneRays|$, for cones even of moderate size, it rapidly becomes infeasible to represent, determine, and work with.  Hence, one often narrows scope to the \emph{affine symmetry group}.
\defn{[Affine Symmetry Group] The affine symmetry group $\affineSymGroup$ consists of only those combinatorial symmetries $\rayPerm$ which can be represented as resulting from an invertible linear transformation from the general linear group $\linearTrans_{\rayPerm} \in \generalLinearGroup$, in the sense that for each extreme ray $\coneRay\in\coneRays$, $\rayPerm(\coneRay) = \linearTrans_{\rayPerm} \coneRay$. }

A further subgroup of the affine symmetry group is the restricted affine symmetry group $\resSymGroup$, can be defined after selecting a series of extreme ray representatives $\coneRayReps$, and replacing each extreme ray $\coneRay = \{\gamma \coneRayRep | \gamma \geq 0 \}$ with its particular selected representative $\coneRayRep$, effectively selecting a scale for this representative.   
\defn{[Restricted Symmetry Group]  The restricted symmetries of a polyhedral cone $\polyCone$ with selected extreme ray representatives $\coneRayReps$ are then those bijections $\rayPerm:\coneRayReps \rightarrow \coneRayReps$ for whom there exists a $\linearTrans_{\rayPerm} \in \generalLinearGroup$ such that $\rayPerm(\coneRayRep) = \linearTrans_{\rayPerm} \coneRayRep$.}  
It is important to note that the same polyhedral cone can thus have multiple restricted symmetry groups via the selection of difficult scalings for the representatives of the extreme rays.

A final, smallest symmetry group for a polyhedron, are the \emph{coordinate permutation symmetries}.

\defn{[Coordinate Permutation Symmetries]  The coordinate permutation symmetry group
 $\permSymGroup(\polyCone)$ of a polyhedral cone, are those affine symmetries $\rayPerm(\coneRay) = \linearTrans_{\rayPerm} \coneRay$ which can be represented by permutation matrices, so that the columns of $\linearTrans_{\rayPerm}$ are the columns of the identity matrix.}

\subsection{Polyhedral Symmetry Groups and the Entropy Region}\label{sec:entSym}
To illustrate these ideas and definitions in an information theoretic context, in this section
we will consider a sequence of examples concerning the entropy cone $\bar{\Gamma}^*_N$ and its outer bound, the Shannon outer bound $\Gamma_N$.

Let's begin by considering the case of the entropy region for two variables $\Gamma^*_2 = \Gamma_2$.  This three dimensional polyhedral cone can be described by three inequalities
\begin{equation}
\Gamma^*_2 = \left\{ \left[\begin{array}{c} h_1 \\ h_2 \\ h_{12} \end{array} \right] \left| \begin{array}{c} h_{12} - h_1 \geq 0 \\ h_{12}- h_2 \geq 0 \\ h_1+h_2 -h_{12} \geq 0 \end{array} \right. \right\}
\end{equation}
and also can be represented as the conic hull of three extreme rays 
\begin{equation}
\Gamma_2^* = \left\{ \left[ \begin{array}{c} 1 \\ 0 \\ 1 \end{array} \right], \left[ \begin{array}{c} 0 \\ 1 \\ 1 \end{array} \right],\left[ \begin{array}{c} 1 \\ 1 \\ 1 \end{array} \right]   \right\}
\end{equation}
Each subset of two of these three extreme rays forms a two dimensional face, so any permutation of the three rays leaves the face lattice invariant, implying that $\combSymGroup(\Gamma^*_2) = \mathbb{S}_3$, the symmetric group of order three.  As there are an equal number of rays as there are dimensions, each of the six ray permutations in $\mathbb{S}_3$ is representable as an invertible matrix transformation, and so the affine symmetry group and restricted affine symmetry group match the combinatorial symmetry group  $\resSymGroup(\Gamma^*_2)=\affineSymGroup(\Gamma^*_2) = \combSymGroup(\Gamma^*_2)$.  The coordinate permutation group, however is a strict subgroup of order 2, which can be viewed as $\permSymGroup(\Gamma_2)=\mathbb{S}_2$, as it contains only the identity and the permutation swapping the coordinates $h_1$ and $h_2$ and leaving $h_{12}$ fixed.

Moving on to $\Gamma_3 = \bar{\Gamma}^*_3$, we have the inequalities
\begin{equation*}
\bar{\Gamma}^*_3=\left\{ \begin{bmatrix} h_1 \\ h_2 \\ h_{12} \\ h_{3} \\ h_{13} \\ h_{23} \\ h_{123} \end{bmatrix} \in \mathbb{R}^7 \left| \begin{array}{c} -h_{12} +h_{123} \geq 0 \\ -h_{13}+h_{123} \geq 0 \\ h_{23}-h_{123} \geq 0 \\ h_1 + h_2 -h_{12} \geq 0 \\ h_1 + h_3 - h_{13} \geq 0 \\ h_2 + h_3 - h_{23} \geq 0 \\ -h_1 + h_{12} + h_{13} - h_{123} \geq 0 \\  
 -h_2 + h_{12} + h_{23} - h_{123} \geq 0 \\
 -h_3 + h_{13} + h_{23} - h_{123} \geq 0
\end{array} \right. \right\}
\end{equation*}
The extreme ray representation $\bar{\Gamma}^*_3=\textrm{conic}(\coneRayReps)$ is
\begin{equation}\label{eq:extrRepShan3}
\coneRayReps:=\left\{
\begin{bmatrix}
     0   \\  1  \\   1  \\   1 \\    1 \\    1  \\   1
\end{bmatrix},
\begin{bmatrix}
     1  \\   0    \\ 1   \\  1    \\ 1  \\   1    \\ 1
\end{bmatrix},
\begin{bmatrix}
     1   \\  1   \\  1   \\  1   \\  1 \\    1  \\   1
\end{bmatrix},
\begin{bmatrix}
     0 \\    1    \\ 1   \\  0    \\ 0    \\ 1    \\ 1
\end{bmatrix},
\begin{bmatrix}
     1 \\    0    \\ 1  \\   0  \\   1  \\   0    \\ 1
\end{bmatrix},
\begin{bmatrix}
     0  \\   0   \\  0  \\   1   \\  1  \\   1   \\  1
\end{bmatrix},
\begin{bmatrix}
     1   \\  1 \\    1  \\   0   \\  1  \\   1   \\  1
\end{bmatrix},
\begin{bmatrix}
     1 \\    1  \\   2   \\  1    \\ 2   \\  2  \\   2
\end{bmatrix}
\right\}
\end{equation}
where we will denote in the same order $\coneRayReps:=\{\coneRayRep_1,\coneRayRep_2,\ldots,\coneRayRep_8\}$.  In addition to the 8 one dimensional faces formed by these rays, and the 1 seven dimensional face formed by all of $\bar{\Gamma}^*_3$ itself, some computation finds 166 other faces, comprised of the numbers of faces of dimensions as indicated in Table \ref{tbl:numFaces}.  Finding the subgroup of $\mathbb{S}_8$ that fixes these faces using \textsf{GAP}, we find the combinatorial symmetry group $\combSymGroup(\bar{\Gamma}^*_3)$ includes 72 permutations generated by the following 5 permutations: swapping $\coneRayRep_1$ with $\coneRayRep_2$, swapping $\coneRayRep_1$ with $\coneRay_7$, swapping $\coneRayRep_5$ with $\coneRayRep_6$, swapping $\coneRayRep_4$ with $\coneRayRep_5$, and swapping $\coneRayRep_3$ with $\coneRayRep_8$.  

In fact, each of the generators for the combinatorial symmetry group is expressible with an invertible linear transformation among the extreme ray representatives depicted in (\ref{eq:extrRepShan3}), which proves that the group of restricted symmetries is in this instance again equivalent to the group of combinatorial symmetries.  Indeed, swapping $\coneRayRep_1$ with $\coneRayRep_2$ and leaving the other rays fixed can be achieved by transforming $h_1 \mapsto h_2+h_{13}-h_{23}=H(1|3)+I(2;3)$ and $h_2 \mapsto h_1 + h_{23} - h_{13} = H(2|3)+I(1;3)$ and leaving all other coordinates unchanged.  Similarly, swapping $\coneRayRep_1$ with $\coneRayRep_7$ and leaving the other rays fixed can be achieved by transforming $h_1 \rightarrow h_3+h_{12}-h_{23} = H(1|2)+I(2;3)$.

Thus, for $\Gamma_3=\bar{\Gamma}^*_3$, like $\Gamma_2=\Gamma_2^*$, the $\combSymGroup(\bar{\Gamma}^*_3)=\affineSymGroup(\Gamma^*_3) =\resSymGroup(\Gamma^*_3)$

The coordinate permutation group $\permSymGroup(\Gamma_3)=\mathbb{S}_3$, as it contains only the identity and the permutation swapping the coordinates $h_1$ and $h_2$ and leaving $h_{12}$ fixed.

\begin{table*}
\centering
\begin{tabular}{cccccccccccccccc}
			&  1 &  2 & 3 & 4 & 5 & 6 & 7 & 8 & 9 & 10 & 11 & 12 & 13 & 14 & 15 \\
$\Gamma_2$ &    3 & 3  &  1 &    &   &    &     &    &    &      &     &      &      &    &  \\
$\Gamma_3$ &    8 & 27& 49& 51& 30& 9& 1 &  &      &     &     &      &     &     & \\
$\Gamma_4$ &   41 & 510 & 3246 & 12654 & 32957 & 60130 & 78868 & 75241 & 52232 & 26112 & 9189 & 2188 & 330 & 28 & 1 \\
\end{tabular}
\caption{Numbers of faces of each dimension for the first few Shannon outer bounds.}\label{tbl:numFaces}
\end{table*}

Moving on to $\Gamma_4 \supsetneq \bar{\Gamma}_4^*$.  In general, the Shannon outer bound $\Gamma_N$ is defined via the $\binom{N}{2}2^{N-2} + N$ elemental inequalities $H(X_i | X_{\mathcal{N}\setminus \{i\}}) \geq 0, i \in\mathcal{N}$ and $I(X_i;X_j | X_{\mathcal{K}}) \geq 0,\ i,j \in \mathcal{N},\ \mathcal{K} \subseteq \mathcal{N}\setminus \{i,j\}, i\neq j$ with $\mathcal{N} =\{1,\ldots,N\}$.  At $N=4$, there are thus 28 such inequalities, and the cone has 41 extreme rays \cite{Hammer2000ShannEntr}, while tedious computation finds the number of faces of each dimension summarized in Table \ref{tbl:numFaces}.  The combinatorial symmetry group $\combSymGroup(\Gamma_4)$ is a subgroup of $\mathbb{S}_{28}$ which fixes this set of faces.  Calculating such a stabilizer subgroup through ordinary routines is daunting for such a large group, so finding the generators and cardinality of $\combSymGroup(\Gamma_N)$ for $N\geq 4$ via a generic such approach appears a daunting task.  However, by leveraging the software \textsf{sympol} \cite{sympol} or the GAP package \textsf{polyhedral} \cite{polyhedralSikiric}, determining the affine symmetry group $\affineSymGroup(\Gamma_4)$ and restricted affine symmetry group $\resSymGroup(\Gamma_4)$ is possible.  By performing this calculation, and a similar one for $\affineSymGroup(\Gamma_5),\affineSymGroup(\Gamma_6),\affineSymGroup(\Gamma_7)$, and studying the structure of these groups with GAP, the following observation can be deduced.

\thml{[Affine Symmetries of the Shannon Outer Bound]  For $N\in\{3,4,5,6,7\}$ the affine symmetry group of the Shannon outer bound, $\affineSymGroup(\Gamma_N)=\resSymGroup(\Gamma_N)$ is isomorphic to $\mathbb{C}_2 \times \mathbb{S}_N \times S_N$, the direct product between the cyclic group of order $2$, the symmetric group of order $N$, and the symmetric group of order $N$.  The normal subgroup isomorphic to $\mathbb{C}_2$ fixes each of the elemental inequalities of the form $H(X_i|X_{\mathcal{N}\setminus i})\geq0$, while swapping $I(X_i;X_j|X_{\mathcal{K}})\geq 0$ with $I(X_i;X_j|X_{(\mathcal{K}\cup\{i,j\})^c})\geq 0$ for each $i,j\in\mathcal{N}$, $i\neq j$ and $\mathcal{K}\subseteq \mathcal{N}\setminus \{i,j\}$.  One normal subgroup isomorphic to $S_N$ permutes the elemental inequalities $H(X_i|X_{\mathcal{N}\setminus i})\geq0$ to $H(X_{\pi(i)}|X_{\mathcal{N}\setminus \{\pi(i)\}})$ for each $\pi \in \mathbb{S}_{\mathcal{N}}$ while leaving all of the other elemental inequalities of the form $I(X_i;X_j|X_{\mathcal{K}})\geq 0$ fixed.  A second normal subgroup leaves each of the elemental inequalities of the form $H(X_i|X_{\mathcal{N}\setminus i})\geq0$ fixed while permuting $I(X_i;X_j|X_{\mathcal{K}})\geq 0$ to $I(X_{\pi(i)};X_{\pi(j)}|X_{\pi(\mathcal{K})}) \geq 0$ for each $\pi \in \mathbb{S}_{\mathcal{N}}$.  Accordingly $|\affineSymGroup(\Gamma_N)|=2(N!)^2$.}{thm:shanSym}

As the coordinate permutation symmetry group $\permSymGroup(\Gamma_N)=\mathbb{S}_N$ is created by random variable permutations, it is clear that the number of affine symmetries of the entropy region is double the square of the number of coordinate permutation symmetries, and is hence far larger.

\subsection{Polyhedral Symmetries in Multiterminal Converse Proofs}\label{sec:netCodSym}
When proving a converse via polyhedral projection for the types of network information theory problems summarized in \S \ref{sec:convProofExamp}, in addition to the symmetries of the entropy cone and its outer bound, the symmetries of the network information theory problem play a role.  In this section, we describe some of these problem symmetries in the context of the type of general network coding problems as described in \S \ref{sec:netCod} and encapsulated in (\ref{eq:netCodRateRegion}), as each of the remaining types of problems described in the remaining subsections of \S \ref{sec:convProofExamp} each have the same form as (\ref{eq:netCodRateRegion}).

In a polyhedral projection problem, there are two polyhedra: the parent polyhedron and the polyhedron resulting from the projection, and both of these can exhibit polyhedral symmetries.   We will observe that both of these types of symmetries can aid in reducing the complexity of computing an explicit rate region.  For projections of the form (\ref{eq:netCodRateRegion}), one can deduce some of the symmetries of the parent polyhedron by restricting the symmetries of the selected outer bound to entropy $\Gamma^o_N$ to those that fix the set of linear constraints $\mathcal{L}_{\mathsf{A}}$, as summarized in the following lemma.

\leml{[Computing Some Network Coding Symmetries]  Subgroups of the symmetries of the parent polyhedron $\Gamma^o_N\cap\mathcal{L}_{\mathsf{A}}$ in (\ref{eq:netCodRateRegion}) can be determined as follows
\begin{eqnarray}
\affineSymGroup(\Gamma^o_N\cap\mathcal{L}_{\mathsf{A}}) &\geq& \textrm{Stabilizer}(\affineSymGroup(\Gamma^o_N),\mathcal{L}_{\mathsf{A}}) \\
\resSymGroup(\Gamma^o_N\cap\mathcal{L}_{\mathsf{A}}) &\geq& \textrm{Stabilizer}(\resSymGroup(\Gamma^o_N),\mathcal{L}_{\mathsf{A}}) \\
\permSymGroup(\Gamma^o_N\cap\mathcal{L}_{\mathsf{A}}) &\geq& \textrm{Stabilizer}(\permSymGroup(\Gamma^o_N),\mathcal{L}_{\mathsf{A}}) 
\end{eqnarray}
where $ \textrm{Stabilizer}$ determines the stabilizer subgroup of the first argument that fixes the second argument, and the groups act on the set of constraints $\mathcal{L}_{\mathsf{A}}$ in the natural way (multiplication of each inequality by an invertible matrix for $\affineSymGroup$ and $\resSymGroup$ and permutation of random variable labels for $\permSymGroup$).
}{lem:someSym}

However, as the following example shows, because they remove both dimensions and inequalities rendered redundant, the equalities in the problem constraints $\mathcal{L}_{\mathsf{A}}$ can in fact yield symmetries beyond those among those of $\Gamma^o_N$, as they remove some of these inequalities, enabling a larger group.

\exmpl{[Network Coding Constraint Symmetry] Substituting the equalities (\ref{eq:ncCons}) associated with the constraints for the network coding problem of Example \ref{ex:run} depicted in Fig. \ref{fig:idsc_ex} into the inequalities describing the entropy cone $\Gamma_6$, one obtains a cone with 264 inequalities, of which 167 can be removed as being redundant after substitution of the equalities (\ref{eq:ncCons}), leaving a cone with 79 inequalities and 54 dimensions associated with the subset entropies remaining after the substitutions from (\ref{eq:ncCons}).  Calculating this cone's restricted symmetry group with the GAP package \textsf{polyhedral} \cite{polyhedralSikiric}, one finds that $\resSymGroup(\Gamma_N \cap \mathcal{L}_{\mathsf{A}})$ is a group with 20 generators and $|\resSymGroup(\Gamma_N \cap \mathcal{L}_{\mathsf{A}})|$ is 358,318,080.  Recalling that $|\resSymGroup(\Gamma_6)|= 2(6!)^2=1,036,800$ it is easily discerned that the substitution of the equalities have removed so many inequalities that $|\resSymGroup(\Gamma_N \cap \mathcal{L}_{\mathsf{A}})|$ is far larger than $\textrm{Stabilizer}(\resSymGroup(\Gamma^o_N),\mathcal{L}_{\mathsf{A}})$.}{ex:bigGroup}

Nonetheless, when the polyhedral cone $\Gamma_N \cap \mathcal{L}_{\mathsf{A}}$ becomes too high dimensional to be amenable to direct computation of symmetry groups, Lem. \ref{lem:someSym} can be used to infer some polyhedral symmetries implied by the problem.  With respect to sets of linear constraints $\mathcal{L}_{\mathsf{A}}$ arising from network coding problems in particular $ \textrm{Stabilizer}(\permSymGroup(\Gamma^o_N),\mathcal{L}_{\mathsf{A}})$ is known as the \emph{network symmetry group} \cite{Li_Operators}.

\defn{[Network Symmetry Group]  The network symmetry group of a network coding rate region problem of the form described in \S \ref{sec:netCod}, is the subgroup of random variable permutations $\mathbb{S}_N$ that fixes the constraint set  $\mathcal{L}_{\mathsf{A}}$ setwise.}

A key benefit of the network symmetry group for network coding rate region problems is that it also forms at least a subgroup of the coordinate symmetry group of the projected rate region $\permSymGroup(\mathcal{R}_o)$.  Indeed, permuting the network coding source and edge rates with the same permutation as their associated random variables leaves the rate region fixed.

\exmpl{[Network Coding Rate Region Symmetry]  The network symmetry group for the problem from Example \ref{ex:run} depicted in Fig. \ref{fig:idsc_ex} is the subgroup of random variable label permutations $\mathbb{S}_{6}$ that fixes the constraints depicted in (\ref{eq:ncCons}).  As will be shown with another example in \S \ref{sec:ITCP}, the GAP package ITCP provided along with the manuscript can calculate this network symmetry group as $\langle (5,6),(4,5) \rangle$, which is a group of order $6$ whose generators swap encoders $5$ and $6$, and encoders $4,5$ independently.  Observe that $\mathcal{R}_o$ in (\ref{eq:epobex}) is fixed under this group, and hence it is a symmetry of both $\mathcal{R}_o$ and the parent polyhedron.  While this group is tiny relative to the restricted symmetry group for this problem uncovered in Example \ref{ex:bigGroup}, even it is already sufficient to demonstrate a substantial reduction in the computational complexity of polyhedral projection in the coming sections.
}{ex:netSymGroup}

\section{Polyhedral Projection with CHM} \label{sec:chm}\label{sec:symCHM}
With the notions of polyhedral symmetries described in \S \ref{sec:polySym}, and the demonstration of their substantial size and non-triviality in network information theory problems provided by the examples in \S \ref{sec:entSym} and \S \ref{sec:netCodSym}, we are ready to turn to the subject of how to exploit symmetry when performing polyhedral projection to determine network information theory rate region as described in \S \ref{sec:convProofExamp}.

The symmetry exploiting polyhedral projection algorithm we will describe builds upon the convex hull method \cite{lassezchm} for projecting polytopes (bounded polyhedra).  As such, we will first review the operation of the convex hull method in this section.  Thereafter, in \S \ref{sec:symchm} we will detail three different ways that known symmetry groups can be exploited to reduce the complexity of the convex hull method, yielding together a new algorithm \textsf{symChm}.

\subsection{Review: The Convex Hull Method}\label{sec:chm}
Denote by $\textsf{proj}_k(\mathcal P)$ the projection of a polyhedron $\mathcal{P}\subseteq \mathbb{R}^d$ onto its first $k$ dimensions
\begin{equation}
\textsf{proj}_k(\mathcal{P})\triangleq \left\{\mathbf{x} \in \mathbb{R}^k \left|  \exists \mathbf{y} \in \mathbb{R}^{d-k}\ \textrm{ s.t. }\ (\mathbf{x},\mathbf{y}) \in \mathcal{P}\right. \right\}
\end{equation}
 CHM \cite{lassezchm} is an algorithm to project polytopes by building successively better inner bounds to the projected polytope $\textsf{proj}_k (\mathcal P)$ via the solution of carefully selected linear programs over $\mathcal P$.  The pseudocode for our implementation  \cite{jayantchm} of CHM is listed as algorithm \ref{algchm}. Note that we assume that the input polyhedron $\mathcal P$ is bounded and full-dimensional, so that the dimension of its affine hull is $d$.  Tests for full-dimensionality and methods for the elimination of any redundant variables and inequalities of the input can be be implemented based on the guidelines in \cite{fukuda2000frequently}.
\begin{algorithm}
\caption{Convex Hull Method}\label{algchm}
\DontPrintSemicolon 
\KwIn{Polyhedron $\mathcal{P}$ and $k$, the dimension of projection}
\KwOut{Vertex-facet set pair $(V,H)$ of $\textsf{proj}_k(\mathcal{P})$}
$(V,H)\leftarrow$  \textsf{initialhull}($\mathcal{P},k$)\;
\While{$\exists \{\mathbf h\mathbf x\geq \mathbf b\}\in H$  s.t. \textsf{isterminal}($\mathcal{P},\mathbf{h},\mathbf{b})= 0$ } {
$\mathbf{v}\leftarrow$ \textsf{extremepoint}($\mathcal{P},\mathbf{h}$)\;
$H\leftarrow\textsf{updatehull}(\mathbf{v},V,H)$\;
$V\leftarrow V\cup{\mathbf{v}}$
}
\Return{$(V,H)$}\;
\end{algorithm}
\begin{procedure}
\DontPrintSemicolon 
\KwIn{Polyhedron $\mathcal{P}$ and $k$, the dimension of projection}
\KwOut{$(V,H)$ corresponding to initial hull of projection}
$V\leftarrow \phi$\;
$\mathbf{p}_1\leftarrow$ \textsf{extremepoint}($\mathcal{P},[1,0,\hdots 0]_{1\times k}$)\;
$\mathbf{p}_2\leftarrow$ \textsf{extremepoint}($\mathcal{P},[-1,0,\hdots 0]_{1\times k}$)\;
$V\leftarrow V\cup\{\mathbf{p}_1,\mathbf{p}_2\}$\;
$\textbf{h}\leftarrow \textsf{hyperplane}(V)$\;\label{proc:inithull_hyperplane}
$i\leftarrow 3$\;
\While{$i\leq k$\label{w1}}{
$\mathbf{p}_i\leftarrow$ \textsf{extremepoint}($\mathcal{P},\mathbf{h}$)\;
$V\leftarrow V\cup\{\mathbf{p}_i\}$\;
$\mathbf{h}\leftarrow$\textsf{hyperplane}($V$)\;
$i\leftarrow i+1$\;\label{w2}
}
$H\leftarrow$ \textsf{facets}($V$)\;\label{proc:inithull_facets}
\Return{$(V,H)$}\;
\label{inithull}
\caption{initialhull($\mathcal{P},k$)}\label{alg:initialHull}
\end{procedure}
\begin{procedure}
\DontPrintSemicolon 
\KwIn{Set of vertices $V$ and corresponding set of inequalities $H$, new vertex $\mathbf{v}$}
\KwOut{Set of inequalities $H'$ corresponding to $V'=V\cup\{\mathbf{v}\}$}
$\mathcal C\leftarrow \textsf{homog}(H)^\circ$\;\label{uph:homog}
$\{\mathbf{r}_1,...,\mathbf{r}_t\}\leftarrow$ \textsf{DDiteration}($\mathcal{C},\{[1 \hspace{2mm} v_1,\hdots, v_d]\mathbf{y}\leq 0\}$)\;
$H'\leftarrow\{\textsf{ineq}(\mathbf r_1),\hdots,\textsf{ineq}(\mathbf r_t)\}$\;
\Return{$H'$}\;
\label{updatehull}
\caption{updatehull($\mathbf{v},V,H$)}
\end{procedure}
\noindent The algorithm relies on the fact that, if $\mathbf{c} \in \mathbb{R}^k$, then a point on the boundary of $\textrm{proj}_k(\mathcal P)$ that attains the solution to the linear program with cost vector $\mathbf{c}$ over $\textrm{proj}_k(\mathcal P)$, can be found by projecting the extreme point in $\mathcal P$ attaining the solution of the linear program with cost vector $[\mathbf{c}^T,\mathbf{0}_{d-k}^T]^T$ over $\mathcal P$, so that
\begin{eqnarray}
\min_{\mathbf{x} \in\textrm{proj}_k \mathcal P} \mathbf{c}^T \mathbf{x}  &=&
\min_{\mathbf{y} \in\mathcal P} [\mathbf{c}^T,\mathbf{0}_{d-k}^T] \mathbf{x}, \ \ \textrm{and}, \\
\arg \min_{\mathbf{x} \in\textrm{proj}_k \mathcal P} \mathbf{c}^T \mathbf{x} &=& \textrm{proj}_k \left( \arg \min_{\mathbf{y} \in\mathcal P} [\mathbf{c}^T,\mathbf{0}_{d-k}^T] \mathbf{x}  \right).
\end{eqnarray}
Furthermore, any extreme point of $\textsf{proj}_k(\mathcal P)$ can be described as a projection of some extreme point of $\mathcal P$, meaning that finding all extreme points of $\textsf{proj}_k(\mathcal P)$ is simply a matter of solving a series of linear programs over $\mathcal P$ and projecting the point attaining the optimum solution.
\subsection{Boundedness Transformation}\label{chm:boundedness}
Note that CHM is made to project polytopes, but when calculating the bounds on network coding rate regions and performing the other types of network information theory calculations described in \S \ref{sec:convProofExamp}, we will be interested in projecting pointed polyhedral cones in $\mathbb R^{d}_{\geq \mathbf 0}$, as all linear information inequalities are homogeneous inequalities while constraints arising from a network coding instance are homogeneous equations.  Fortunately, an unbounded polyhedron $\mathcal C$ can be transformed to create a polytope $\mathcal{B}(\mathcal C)$ such that the projection of the unbounded polyhedron $\mathcal C$ can be obtained from projection of $\mathcal B(C)$.  While there are several such transformations, including the one in \cite{lassezchm}, we describe a transformation that is more efficient in terms of dimension of $\mathcal B(\mathcal C)$ (=$d$).  For an arbitrary polyhedron in $\mathcal P\subseteq \mathbb R^{d}_{\geq \mathbf 0}$, one can apply the transformations described in this section to $\textsf{homog}(\mathcal P)\subseteq \mathbb R^{d+1}_{\geq \mathbf 0}$. Let $\mathbf{Hx}\geq \mathbf 0$ be the inequality description associated with a polyhedral cone $\mathcal C$.  This cone can be transformed into a polytope
$
\mathcal C'=\{\mathbf{x} \in \mathbb R^{d}_{\geq \mathbf 0}| \mathbf{Hx}\geq \mathbf 0\ \wedge\  (\mathbf{1}_k^T,\mathbf{0}_{d-k}^t )\mathbf{x}\leq 1\}
$.
\leml{
The inequality representation of $\textsf{proj}_k(\mathcal C')$ is equal to the inequality description of $\textsf{proj}_k(\mathcal C)$ plus the additional inequality $\mathbf{1}_k^T\mathbf x_{1:k}\leq 1$.
}{lem:bounded1}
\noindent \textbf{Proof}:  The inequalities in the inequality representation of $\textsf{proj}_k(\mathcal C)$ must be the only homogeneous inequalities in the representation of $\mathcal C'$, as adding a non-homogeneous inequality $[\mathbf{1}_k^T,\mathbf 0_{d-k}^t ]\mathbf x\leq 1$ to $\mathcal C$ cannot affect their minimality. Next, we must show that $\mathbf{1}_k^T\mathbf x_{1:k}\leq 1$ is the only non-homogeneous inequality that is non-redundant for $\textsf{proj}_k(\mathcal C')$. Assuming the contrary, there must be an inequality $\mathbf a^T\mathbf x_{1:k}\leq b$ that is also non-redundant for $\textsf{proj}_k(\mathcal C')$. Every inequality bounding  $\textsf{proj}_k(\mathcal C')$ can be obtained as a conic combination of inequalities in the inequality representation of $\mathcal C'$. Hence, $(b,\mathbf a^T)=\boldsymbol\lambda \mathbf H+\lambda_0(1,\mathbf 1_{k}^T)$, where $\boldsymbol\lambda$ is a non-negative real vector and $\lambda_0$ is a non-negative real number. Furthermore, since $(\mathbf{1}_k^T,\mathbf 0_{d-k}^t )\mathbf x\leq 1$ is the only non-homogeneous inequality in the inequality representation of $\mathcal C'$, $\lambda_0\neq 0$. Now, $\boldsymbol\lambda \mathbf H\geq 0$ is a homogeneous inequality satisfied by projection, and hence can be written as a conic combination of non-redundant homogeneous inequalities in the  inequality representation of $\textsf{proj}_k(\mathcal C')$. This means, $\mathbf a^T\mathbf x\leq b$ can be written as a conic combination of inequalities bounding $\textsf{proj}_k(\mathcal C)$ and $\mathbf{1}_k^T\mathbf x_{1:k}\leq 1$, contradicting our assumption that it is non-redundant w.r.t. $\textsf{proj}_k(\mathcal C')$. $\blacksquare$

Using lemma \ref{lem:bounded1}, given the inequality description of $\textsf{proj}_k(\mathcal C')$, we can obtain the inequality representation of $\textsf{proj}_k(\mathcal C)$ by simply deleting the only non-homogeneous inequality present in the representation.

\exmp{[Boundedness Transformation for Ex. \ref{ex:run}] In context of the IDSC instance in Fig. \ref{fig:idsc_ex}, the bounding inequality is of the form
\begin{equation}
\omega_1+\omega_2+\omega_3+R_4+R_5+R_6\leq 1 .
\end{equation}
}

\subsection{Convex Hull Method: low-level details}\label{polyhedra:chminithull}
The first step in CHM, encapsulated in procedure \textsf{initialhull}, is the construction of an inner bound to $\textsf{proj}_k(\mathcal P)$, which is the convex hull of $d+1$ points on the boundary of $\textsf{proj}_k(\mathcal P)$, that is itself full-dimensional. In order to obtain the initial inner bound for the process, we first obtain two boundary points of $\textrm{proj}_k \mathcal P$ by solving two linear programs,
$\min_{\mathbf{y}\in\mathcal P} [-1,\mathbf{0}_{d-1}^T] \mathbf{y}$ and $\min_{\mathbf{y}\in\mathcal P} [1,\mathbf{0}_{d-1}^T] \mathbf{y}$.  We  then select the normal vector $\mathbf c$ of any hyperplane containing these points (proc. $\textsf{hyperplane}$), and new boundary points of the projection are obtained as $\textrm{proj}_{k} \arg\min_{\mathbf{y}\in\mathcal P} [\mathbf{c},\mathbf{0}_{d-k}] \mathbf{y}$ and $\textrm{proj}_k \arg\min_{\mathbf{y}\in\mathcal P} [-\mathbf{c},\mathbf{0}_{d-k}] \mathbf{y}$. This process of finding a hyperplane containing all previously known boundary points and finding new boundary points is repeated until we obtain $k+1$ boundary points of the projection that give a full dimensional initial inner bound of the projection (in other words, until we obtain $k+1$ convex-independent boundary points in $\mathbb R^k$). This process needs to be repeated at most $k+1$ times, owing to the full-dimensionality of $\mathcal P$. Given the set $V$ containing $k+1$ boundary points of the projection forming a full-dimensional inner bound, computing the associated inequality description $\textsf{conv}(V)$ corresponds to a $k+1\times k+1$ matrix inversion (proc. $\textsf{facets}$). This gives us a double description pair of the first full dimensional inner bound of the projection.

\exmp{[Initial Hull for Ex. \ref{ex:run}] We now describe the construction of initial hull in the context of IDSC instance in Fig. \ref{fig:idsc_ex}, which in this case is the polytope $\mathcal B_1\triangleq\textsf{conv}(\{\mathbf v_i| i\in \{1,\ldots,7\}\})$. The vertices of $\mathcal B_1$ are obtained by solving a series of linear programs with carefully chosen objective functions.  The summary of the $\textsf{initialhull}(\cdot)$ procedure can be found in table \ref{tab:inithull_v}. The first and second Linear programs have objective functions $\omega_1$ and $-\omega_1$ respectively. At any later step $i>2$, the normal vector (denoted as $\mathbf h$ in $\textsf{initialhull}(\cdot)$ line \ref{proc:inithull_hyperplane}) of any of the several hyperplanes containing vertices $\mathbf v_1,\hdots,\mathbf v_{i-1}$ is computed using procedure $\textsf{hyperplane}(\cdot)$. The objective function is then $\mathbf h$ itself or its negation $-\mathbf h$. The assumption, that the projection polytope is full-dimensional, implies that at least one of these two linear programs must yield a vertex that is convex independent w.r.t. the vertices found so far. Once, we have found $k+1=7$ convex independent vertices, the inequality description of the convex hull of $\mathbf v_1,\hdots,\mathbf v_7$ is obtained using the $\textsf{facets}(\cdot)$ procedure (line \ref{proc:inithull_facets} of $\textsf{initialhull}(\cdot)$). Internally, the procedure $\textsf{facets}(\cdot)$ constructs the inequality description by inverting the matrix $\mathbf A_1$ associated with the inequality description of the cone $\mathcal P_1\triangleq\mathcal C(\mathcal B_1)^\circ$. The inequality description of $\mathcal P_1$ is $\mathcal P_1(\mathbf A_1,\mathbf 0)$ where,
\begin{equation}
\mathbf A_1 \triangleq \begin{bmatrix}
-1 & -\frac{1}{2} & 0 & 0 & 0 & -\frac{1}{2} & 0\\
-1 & 0 & 0 & 0 & 0 & 0 & 0\\
-1 & -\frac{1}{2} & 0 & 0 & 0 & 0 & -\frac{1}{2}\\
-1 & 0 & 0 & 0 & 0 & -1 & 0\\
-1 & 0 & -\frac{1}{4} & 0 & -\frac{1}{4} & -\frac{1}{4} & -\frac{1}{4}\\
-1& 0 & 0 & 0 & -1 & 0 & 0 \\
-1 & 0 & 0 & -\frac{2}{5} & \frac{1}{5} & \frac{1}{5} & \frac{1}{5}\\
\end{bmatrix}
\end{equation}
The extreme rays of $\mathcal P_1$, obtained by computing $\mathbf A_1^{-1}$, are given by the columns of matrix $\mathbf Y$ where,
\begin{equation}
\mathbf Y\triangleq
\begin{bmatrix}
0 & 1 & 0 & 0 & 0 & 0 & 0\\
1 & -1 &  0 &  -1 &  0 &  0 &  0\\
1 &  -1 &  -1  &  -2 &  1  & -1 &  0\\
\frac{1}{2} &  -1 &  -\frac{1}{2} &  -1 &  0 &  -\frac{1}{2} &  2\\
0 &  -1 &  0 &  0 &  0 &  1 &  0\\
0 &  -1 &  0 &  1 &  0 &  0 &  0\\
-1 &  -1 &  1 &  1 &  0 &  0 &  0\\
\end{bmatrix}
\end{equation}
The inequality description of $\mathcal B_1$ can be then obtained by reading off the columns of matrix $\mathbf Y$ as,
\begin{equation}\label{eq:inithull_h}
\begin{aligned}
2 R_6 &\leq 2\omega_1+2\omega_2+\omega_3 & (\mathbf h_1) \\
\omega_1 +\omega_2 + \omega_3 + R_4 + R_5 + R_6 &\leq 1& (\mathbf h_2)\\
2\omega_2+\omega_3 &\leq 2 R_6 & (\mathbf h_3)\\
\omega_1+2\omega_2+\omega_3 &\leq R_5+R_6 & (\mathbf h_4)\\
-\omega_2 &\leq 0 & (\mathbf h_5)\\
2\omega_2 +\omega_3 &\leq 2R_4 & (\mathbf h_6)\\
 -\omega_3 &\leq 0 & (\mathbf h_7)
\end{aligned}
\end{equation}
}
\begin{table*}
\vspace{3mm}
\begin{center}
\begin{tabular}{|c|c|c|c|}
\hline
LP & objective function & new vertex & homogenized vertex\\
\hline\hline
1 & $\omega_1$ & $\mathbf{v}_1=(\frac{1}{2},0,0,0,\frac{1}{2},0)$ & $\mathbf{r}_1=(\frac{1}{2},0,0,0,\frac{1}{2},0)$  \\
\hline
2 & $-\omega_1$ & $\mathbf{v}_2=(0,0,0,0,0,0)$ & $\mathbf{r}_2=(0,0,0,0,0,0)$ \\
\hline
3 & $\omega_1-R_2$ &  $\mathbf{v}_3=(\frac{1}{2},0,0,0,0,\frac{1}{2})$ &  $\mathbf{r}_3=(\frac{1}{2},0,0,0,0,\frac{1}{2})$ \\
\hline
4 & $-\omega_1+R_2$ & $\mathbf{v}_4=(0,0,0,0,1,0)$ & $\mathbf{r}_4=(0,0,0,0,1,0)$ \\
\hline
5 & $\omega_2$ & $\mathbf{v}_5=(0,\frac{1}{4},0,\frac{1}{4},\frac{1}{4},\frac{1}{4})$ & $\mathbf{r}_5=(0,\frac{1}{4},0,\frac{1}{4},\frac{1}{4},\frac{1}{4})$ \\
\hline
6 & $-\omega_2$ & $-$ & $-$\\
\hline
7 & $\omega_2-R_1$ & $-$ & $-$\\
\hline
8 & $-\omega_2+R_1$ & $\mathbf{v}_6=(0,0,0,1,0,0)$ & $\mathbf{r}_6=(0,0,0,1,0,0)$ \\
\hline
9 & $\omega_3$ & $\mathbf{v}_7=(0,0,\frac{2}{5},\frac{1}{5},\frac{1}{5},\frac{1}{5})$ & $\mathbf{r}_7=(0,0,\frac{2}{5},\frac{1}{5},\frac{1}{5},\frac{1}{5})$ \\
\hline
\end{tabular}
\vspace{3mm}
\caption{Summary of the linear programs solved during the $\textsf{initialhull}(\cdot)$ procedure in CHM.}\label{tab:inithull_v}
\end{center}
\end{table*}

Once the initial hull is computed, at each stage of the CHM algorithm, a DD pair is maintained for the current inner bound to $\textrm{proj}_k \mathcal P$.  Each inequality in the inequality description of the inner bound carries with it a label, indicating whether or not it is terminal or non-terminal.  The initial inner bound has all of its inequalities labelled as non-terminal.  A non-terminal inequality $\mathbf{c}^T \mathbf{x} \geq b$ is then selected (proc. \textsf{isterminal}), and the linear program $\min_{\mathbf{x} \in \mathcal P} [\mathbf{c}^T\mathbf{0}_{d-k}^T] \mathbf{x} $ is solved over the high dimensional polyhedron.  If the solution obtained is $b$, the inequality is marked as terminal.  Otherwise, the extreme point of $\mathcal P$, obtained in the process, i.e. $\mathbf{v}=\arg\min_{\mathbf{x} \in \mathcal P} [\mathbf{c}^T\mathbf{0}_{d-k}^T] \mathbf{x} $ attaining the solution is projected to get a new boundary point $\textrm{proj}_k \mathbf{v}$ of $\textrm{proj}_k \mathcal P$.  The procedure \textsf{extremepoint}($\mathcal{P},\mathbf{c}$) thus returns $\textrm{proj}_k \left( \arg \min_{\mathbf{y} \in\mathcal P} [\mathbf{c}^T,\mathbf{0}_{d-k}^T] \mathbf{x}  \right)$.  The DD pair of the inner bound is then updated (proc. \textsf{updatehull}) by adding the new boundary point, viewing it as a new inequality in the polar cone, via a single DD algorithm update step \cite{fukuda1996double} (proc. \textsf{DDiteration}), and any new inequalities thus introduced are marked as non-terminal.  In this double description iteration, the new extreme point (ray of homogenization) is viewed as a new inequality for the polar of this homogenization, and the resulting new extreme rays are, by passing again back through the polar, are thus directly interpreted (by procedure \textsf{ineq}) as new inequalities of the inner bound created by adding this ray.  Finally, a new non-terminal facet is selected and the process is repeated until all of the facets are labelled as terminal, at which point the inner bound has been proven equal to $\textrm{proj}_k \mathcal P$.

To illustrate these remaining iterations of CHM after the initial hull is computed, the next example provides a summary of the remaining iterations for the case of the running example.
\exmp{[CHM for Ex. \ref{ex:run}] Table \ref{tab:chmsummary} gives a summary of all linear programs solved, new vertices revealed, and non-facets revealed after initial hull is computed in the context of IDSC instance in Fig. \ref{fig:idsc_ex}.}
\begin{table*}
\begin{center}
\begin{tabular}{|c|c|c|c|c|}
\hline
LP & objective function (facet) & new vertex & non-facets & new facets\\
\hline\hline
10 & $-2\omega_1-2\omega_2-\omega_3+2R_6$ ($\mathbf h_1$) & $\mathbf{v}_8=(0,0,0,0,0,1)$ & $\mathbf h_1$ &  $\begin{aligned} -\omega_1 & \leq 0 & (\mathbf h_9)\\ 2\omega_2+\omega_3 &\leq 2R_5 & (\mathbf h_{10})\end{aligned}$ \\
\hline
11 & $2\omega_2+\omega_3-2R_6$ ($\mathbf h_{3}$) & $\mathbf v_9=(0,0,\frac{1}{3},\frac{1}{3},\frac{1}{3},0)$ & $\mathbf h_{3}$ & $\begin{aligned} \omega_2 &\leq R_6 & (\mathbf h_{11})\\ 2\omega_2+\omega_3 & \leq R_4+R_6 & (\mathbf h_{12}) \end{aligned}$\\
\hline
12 & $2\omega_2+\omega_3-2R_4$ ($\mathbf h_{6}$) & $\mathbf v_{10}=(0,0,\frac{1}{3},0,\frac{1}{3},\frac{1}{3})$ & $\mathbf h_{6}$ & $\begin{aligned} \omega_2 &\leq R_4 & (\mathbf h_{13})\\ 2\omega_2+\omega_3 &\leq R_4 + R_5 & (\mathbf h_{14})\\ \omega_1+4\omega_2+2\omega_2 &\leq 2R_4+R_5+R_6 & (\mathbf h_{15}) \end{aligned}$\\
\hline
13 & $-\omega_3$ ($\mathbf h_7$)  & - & - & - \\
\hline
14 & $2\omega_2+\omega_3-R_5$ $(\mathbf h_{10})$ & $\mathbf v_{11}=(0,0,\frac{1}{3},\frac{1}{3},0,\frac{1}{3})$ & $\mathbf h_{10}$ & $\begin{aligned} \omega_2 &\leq R_5 & (\mathbf h_{16})\end{aligned}$\\\hline

15 & $\omega_2-R_5$ ($\mathbf h_{16}$)  & - & - & - \\ \hline
16 & $2\omega_2+\omega_3-R_4-R_6$ ($\mathbf h_{12}$)  & - & - & - \\  \hline
17 & $\omega_3-R_4$ ($\mathbf h_{13}$)  & - & - & - \\  \hline
18 & $\omega_1+4\omega_2+2\omega_3-2R_4-R_5-R_6$ ($\mathbf h_7$)  & - & - & - \\ \hline
19 & $\omega_1+2\omega_2+\omega_3-R_5-R_6$ ($\mathbf h_{4}$) & $\mathbf v_{12}=(\frac{1}{2},0,0,\frac{1}{2},0,0)$ & $\mathbf h_{4}$ & $\begin{aligned} 2\omega_2+\omega_3 &\leq R_5+R_6 & (\mathbf h_{17})\\
2\omega_1+6\omega_2+3\omega_3 &\leq 2R4+2R_5+2R_6 & (\mathbf h _{18}) \\
\omega_1+4\omega_2+2\omega_3 &\leq R4+2R_5+R_6 & (\mathbf h _{19}) \\
\omega_1+4\omega_2+2\omega_3 &\leq R4+R_5+2R_6 & (\mathbf h _{20})
\end{aligned}$\\\hline
20 & $-\omega_1$ ($\mathbf h_{9}$)  & - & - & - \\ \hline
21 & $2\omega_2+\omega_3-R_4-R_5$ ($\mathbf h_{14}$)  & - & - & - \\  \hline
22 & $\omega_1+\omega_2+\omega_3-R_4-R_5-R_6$ ($\mathbf h_2$)  & - & - & - \\ \hline
23 & $-\omega_2$ ($\mathbf h_{16}$)  & - & - & - \\ \hline
24 & $\omega_2-R_5$ ($\mathbf h_{16}$)  & - & - & - \\  \hline
25 & $\omega_1+4\omega_2+2\omega_3 - R4-R_5-2R_6$ ($\mathbf h_{20}$)  & - & - & - \\  \hline
26 & $2\omega_1+6\omega_2+3\omega_3-2R_4-2R_5-2R_6$ ($\mathbf h_{18}$)  & - & - & - \\ \hline
27 & $2\omega_2+\omega_3-R_5-R_6$ ($\mathbf h_{17}$)  & - & - & - \\ \hline
28 & $\omega_1+4\omega_2+2\omega_3 -\leq R4-2R_5-R_6$ ($\mathbf h_{19}$)  & - & - & - \\  \hline
\end{tabular}
\vspace{3mm}
\caption{Summary of the linear programs solved after the $\textsf{initialhull}(\cdot)$ procedure in CHM.}\label{tab:chmsummary}
\end{center}
\end{table*}

These remaining iterations also make use of the $\textsf{updatehull}(\cdot)$ procedure, involving a double description set, and the next example demonstrate how this procedure works for the first update after the initial hull for Example \ref{ex:run}.

\begin{figure}[h]
\begin{center}
\includegraphics[scale=1.2]{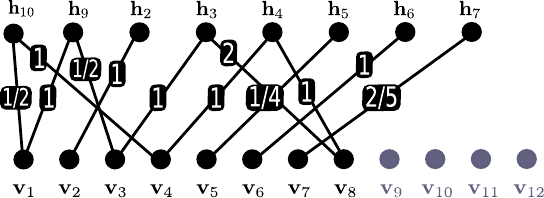}
\end{center}
\caption{Complement of the vertex-facet incidence graph of inner bound $\mathcal B_2$, produced at the end of procedure $\textsf{updatehull}(\cdot)$ after augmenting $\mathcal B_1$ with newly revealed vertex $\mathbf v_8$.}\label{fig:ddex_c}
\end{figure}

\exmp{[$\textsf{updatehull}$ immediately after \textsf{initialhull}, Ex. \ref{ex:run}] 
The input to the $\textsf{updatehull}(\cdot)$ procedure is the inequalities (see \eqref{eq:inithull_h}) and vertices (see table \ref{tab:inithull_v}) of $\mathcal B_1$, along with the newly revealed vertex $\mathbf v_8$. Since $\mathbf v_8$ was obtained as the optimum vertex  of the linear program with cost function $-2\omega_1-2\omega_2-\omega_3+2R_6$, it violates inequality $\mathbf h_1$. In this case, it is the only inequality in the inequality description of $\mathcal B_1$ violated by $\mathbf v_8$. Thus, the inequality description of $\mathcal B_2\triangleq \textsf{conv}(\{\mathbf v_i| i\in \{1,\ldots,8\}\})$ must contain all inequalities in the inequality description of $\mathcal B_1$ except $\mathbf h_1$. As for the new inequalities that hold for $\mathcal B_2$, the original CHM algorithm proposed by Lassez et. al. would resort to a combinatorial search, where the search space would be all $6$-subsets of $\{\mathbf v_i| i\in \{1,\ldots,8\}\}$, any of which define a unique supporting halfspace for $\mathcal B_2$, provided it contains $\mathbf v_8$ . The authors' implementation of CHM, instead reduces the problem of updating $\mathcal B_1$ with $\mathbf v_8$ to that of intersecting $\mathcal C(\mathcal B_1)^\circ$ with a halfspace obtained by putting $\mathbf v_8$ through homogenization and polar transformation. To avoid confusion, we present a description of this procedure, in the context of $\mathcal B_1$ itself, as opposed to $\mathcal C(\mathcal B_1)^\circ$ (which can be thought of as primal vs dual interpretations of the same procedure, with duality in question being the polar duality).

In general, let $J^+$ and $J^-$ be the sets of inequalities that strictly satisfy and violate the newly revealed vertex respectively. The main lemma of the double description method (see \cite{fukuda1996double}, lemma 3) dictates that it suffices to consider conic combinations of pairs of inequalities in $J^+\times J^-$ to construct the new inequalities of the inequality description of the augmented inner bound. Fig. \ref{fig:ddex_b} describes the incidence relationships between newly revealed vertex $\mathbf v_8$ and the facets of $\mathcal B_1$. In this case, $J^+=\{\mathbf h_1\}$ and $J^-=\{\mathbf h_3,\mathbf h_4\}$. Furthermore, the strengthened main lemma of the double description method (see \cite{fukuda1996double}, lemma 8), shows that it suffices to restrict attention to the pairs $(\mathbf h_i,\mathbf h_j)\in J^+\times J^-$ that are adjacent i.e. the intersection of the associated facets have dimension $k-2$ where $k$ is the dimension of the polytope in question. Fukuda et al. \cite{fukuda1996double} also provide simple tests of adjacency, and a technique for computing conic combination coefficients to be used to combine the inequalities, which we omit for the sake of conciseness. In our example, every pair of inequalities in $J^+\times J^-$ is adjacent, and they can be combined as follows to obtain new inequalities $\mathbf h_9$ and $\mathbf h_{10}$.
\begin{equation}
\begin{aligned}
-2\omega_1 -& 2\omega_2 &-\omega_3 &+R_6&\leq 0 & (\mathbf h_1)\\
+& 2\omega_2 &+\omega_3 &-R_6 &\leq 0 & (\mathbf h_3)\\
 \cline{1-6}
-2\omega_1 & & & & \leq 0 & (\mathbf h_9)
\end{aligned}
\end{equation}
\begin{equation}
\begin{aligned}
-2\omega_1 &- 2\omega_2 &-\omega_3 & &+R_6&\leq 0 & (\mathbf h_1)\\
+\omega_1 &+ 2\omega_2 &+\omega_3 &-R_5 &-R_6 &\leq 0 & (\mathbf h_4)\\
 \cline{1-7}
&\omega_2 &+\frac{1}{2}\omega_3 &-R_5 & & \leq 0 & (\mathbf h_{10})
\end{aligned}
\end{equation}
In addition to adding these new inequalities, the routine \textsf{updatehull}, updates the vertex facet incidence relationships for the new inner bound $\mathcal{B}_2$, to those shown in Fig. \ref{fig:ddex_c}.
}

\begin{figure}[h]
\begin{center}
\includegraphics[scale=1.2]{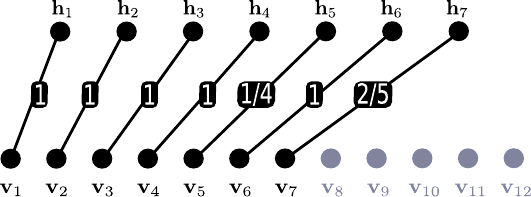}
\end{center}
\caption{Complement of the vertex-facet incidence graph of inner bound $\mathcal B_1$, produced at the end of procedure $\textsf{initialhull}(\cdot)$. The nodes on top represent the facets (see \eqref{eq:inithull_h}), while the nodes at the bottom represent the vertices of $\mathcal B_1$ (see table \ref{tab:inithull_v}). The grayed-out vertices lie outside $\mathcal B_1$, and are yet undiscovered. The numbers on the edges give the strictness with which an inequality is satisfied. }\label{fig:ddex_a}
\end{figure}

\begin{figure}[h]
\begin{center}
\includegraphics[scale=1.2]{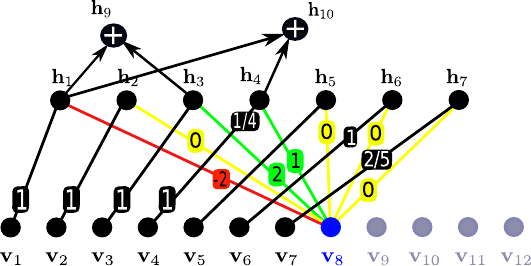}
\end{center}
\caption{The graph in Fig. \ref{fig:ddex_a}, superimposed with incidence relationship between facets $\mathbf h_1,\hdots,\mathbf h_7$ and the newly revealed vertex $\mathbf v_8$. The colors red, yellow and green represent violation, equality, or strict satisfaction of a particular inequality by $\mathbf v_8$. The inequalities violated by $\mathbf v_8$ ($\mathbf h_1$) are combined with the inequalities strictly satisfied by $\mathbf v_8$ ($\mathbf h_3,\mathbf h_4$) to produce the new inequalities that hold for the augmented inner bound ($\mathbf h_9,\mathbf h_{10}$). }\label{fig:ddex_b}
\end{figure}

\subsection{CHM Complexity: Worst Case and a Series of Examples}
In this subsection, we consider a simple family of HMSNC instances and gauge the practical performance of CHM for computation of the associated EPOBs, along with a discussion of worst case complexity of CHM and its various components. When computing an EPOB for a HMSNC instance with $N$ random variables, with $\Gamma_{\textsf{out}}=\Gamma_N$, one starts with the polytope defined by the inequality representation containing elemental Shannon-type inequalities \cite{yeungframework}, network constraints and the bounding inequality $\sum_{i\in[k]}\boldsymbol \omega_i+\sum_{j\in[N]\setminus [k]} R_j\leq 1$. This polytope is then projected to $\boldsymbol \omega,\mathbf r$ coordinates using CHM and from the resultant inequality representation we simply remove the bounding inequality, giving us the inequality representation of the EPOB.
\begin{figure}[h]
\begin{center}
\includegraphics[width=3in]{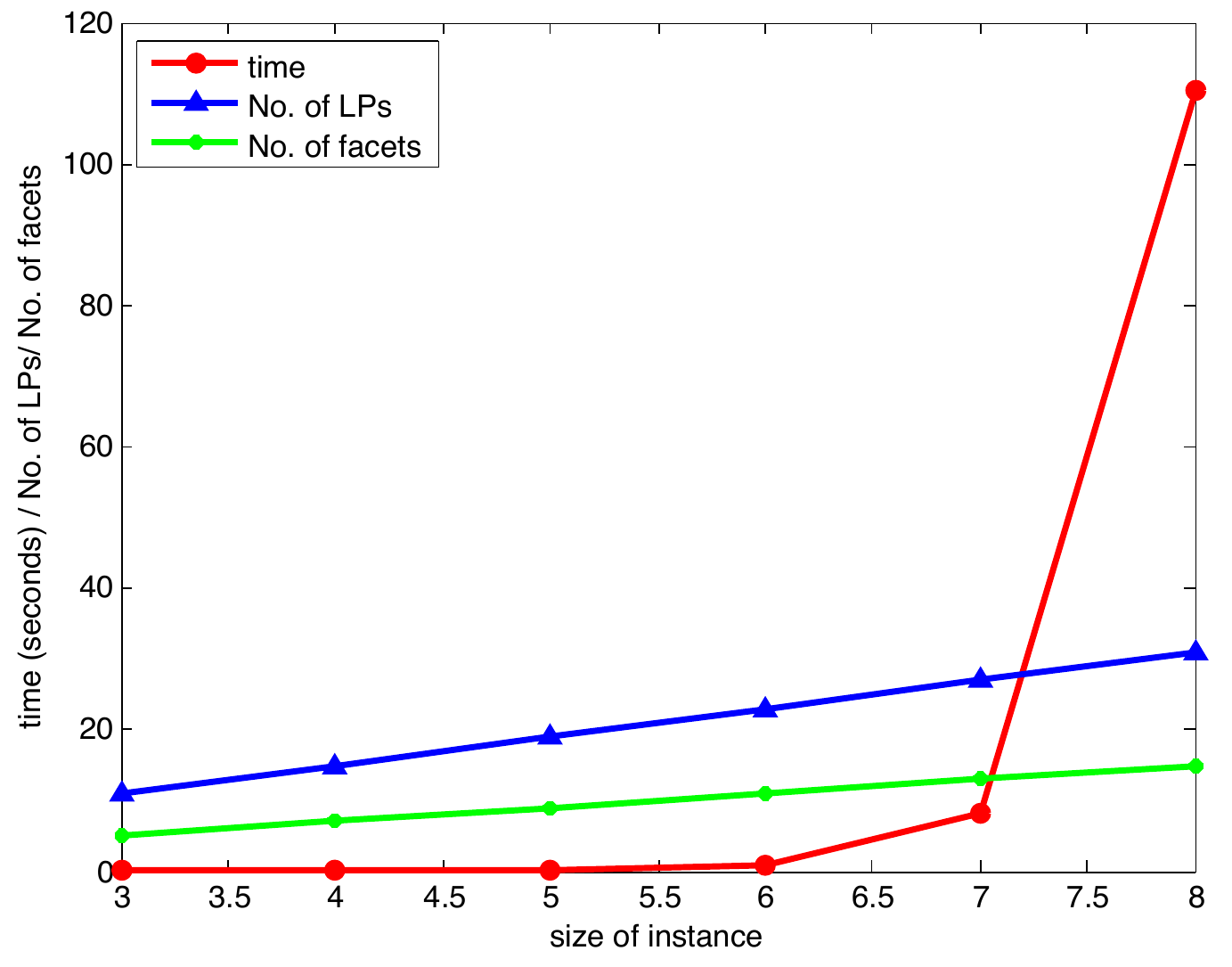}
\end{center}
\caption{The runtime, number linear programs solved, and no. of facets of rate region outer bound, for computing EPOB for a $U^2_k$ network vs $k$, the size of network, with $\Gamma_{\textsf{out}}=\Gamma_k$ i.e. the Shannon outer bound.}\label{fig:u2k}
\end{figure}

\exmpl{$U^2_k$ networks \cite{zongpenguniform} are a family of matroidal networks, that can be constructed from uniform matroids of rank $2$ on ground set of size $k$, using the construction of Dougherty, Freiling and Zeger\cite{DFZMatroidNetworks}. Fig. \ref{fig:u2k} shows the plot of no. of linear programs solved by CHM and time taken by CHM as a function of $k$, which is the size of a $U^2_k$ network. The number of linear programs solved by CHM, which in this case, is the sum of number of facets and number of vertices of the EPOB, grows linearly w.r.t. $k$. On the other hand, the runtime grows exponentially w.r.t. $k$. The growth in runtime reflects the exponential growth in the dimension of the polyhedra being projected, as the outer bound $\Gamma_k$, which is used to compute these EPOBs, lives in a $2^k-1$ dimensional space.
}{u2kexmp}

Now we point out worst case complexities of some of the building blocks of CHM, which are daunting, in keeping with the worst case complexities of most polyhedral computation algorithms. Note that the rate region outer bounds computed in example \ref{u2kexmp} are very \textit{well-behaved} when compared to McMullen's upper bound of $\mathcal O(m^{\left\lfloor\frac{k}{2}\right\rfloor})$\cite{ziegler1995lectures} on the no. of vertices of a $k$ dimensional polytope specified by $m$ inequalities, as are many polyhedra.  It is possible, and in fact highly likely based on experience with examples, that the EPOBs for HMSNC instances and other network information theory problems satisfy a tighter worst case bound on the number of extreme rays when $\Gamma_{\textsf{out}}=\Gamma_N$.  This is related to the enumeration of extremal polymatroids, i.e. extreme rays of the cone $\Gamma_N$, which remains an open problem, although it is lower bounded by the number of connected matroids \cite{Nguyen1978,Li_Allerton_13,Mayhew2008415}, which has been more carefully studied.  

Owing the capability to generate severely degenerate polyhedra having many bases per extreme point, the simplex method, which is used in CHM to solve linear programs exactly by employing rational arithmetic, in worst case complexity analysis can require exponentially many pivots in $d$ for all known pivoting methods. Hence, in the worst case, assuming that the only points of projection obtained during the course are the extreme points of the projection, CHM could solve  $\mathcal O(m^{\left\lfloor\frac{k}{2}\right\rfloor})+m$ linear programs, one per vertex and facet of projection, with each having exponential runtime, given that the projection has $m$ facets. Furthermore, the double description method, used in CHM to incrementally build the projection polytopes, does not have an output sensitive time complexity \cite{Bremner1996convhull}, as it is known to be susceptible to the insertion order of inequalities. In the case of CHM, this means that the order in which we find the extreme points of the projection, greatly affects the number of rays in intermediate double description pairs. Every iteration of DD method, the number of extreme rays can get squared, in the worst case. 

On the other hand, compared with Fourier-Motzkin, the most commonly taught method for small polyhedral projections, CHM can have vastly improved computational complexity.  In Fourier-Motzkin, one calculates not only the projected final polyhedron, but also every intermediate projected polyhedron of intermediate dimension between that of the parent polyhedron.  These polyhedra of intermediate dimension typically have far larger numbers of inequalities and extreme rays than those of the parent or projected polyhedron, hence for problems such as the network coding rate regions, non-Shannon inequalities, linear rank inequalities, etc. in which the number of dimensions being removed is exponential in the number of dimensions being kept, algorithms such as CHM which work directly and exclusively in the projected polyhedron by solving linear programs have substantially lower complexity than Fourier-Motzkin.

While the worst case analysis of the building blocks of CHM paints an ominous picture, both simplex method and double description method have been practically successful in solving moderate to large problem instances, which is essentially what the authors expect of CHM, and what Ex. \ref{u2kexmp} and nearly one million other worked examples \cite{Li_Operators} demonstrates.  

As solving linear programs and computing double descriptions updates are inherent to algorithms aiming to reduce complexity relative to Fourier Motzkin by working directly in the projected space, it is typically in the polyhedral computation literature to assess their complexity in terms of numbers of linear programs that must be solved, the dimensions and numbers of constraining inequalities in these linear programs, and the number of double descriptions steps that must be performed \cite{FukudaPolyCompNotes}.  As each step of CHM solves a linear program which reveals a new extreme point of the projected region, it is clear that the number of linear programs which must be solved is equal to the number of extreme points of the projected region, with each such linear program having a dimension and number of inequalities consistent with the parent polyhedron.  Similarly, CHM must compute an equal number of double descriptions steps.  Bearing in mind that solving linear programs with simplex and computing double descriptions steps can in worst case have high complexity, but are routinely performed in many contexts, and in fact are necessary for any projection algorithm working exclusively in the projection dimensions, in the next section we shift our attention further reducing the complexity of CHM by decreasing the number of linear programs, their dimensions, and double descriptions steps that must be solved. 

\section{A Polyhedral Projection Algorithm that Exploits Symmetry}\label{sec:symchm}
Having assembled the necessary ingredients in previous sections, in this section we are prepared to present the main contribution of the manuscript: a polyhedral projection algorithm that can exploit known symmetry groups of the projected polyhedron and the parent polyhedron, and the demonstration of its complexity reduction benefits when computing network information theory rate regions.

The projection algorithm we have created, named symCHM, whose pseudocode is presented in \ref{symchm}, exploits symmetry to reduce the complexity of CHM in three different ways.  Additionally, substantial storage/memory reduction is achieved.  These reductions build from symmetry exploiting techniques in the polyhedral computation literature.
Bremner et. al. \cite{Bremner09polyhedralrepresentation} and Rehn \cite{RehnSymmetries} consider the problem of exploiting the knowledge of restricted affine symmetries in polyhedral representation conversion, which is the problem of computing the inequality representation of a polyhedron, given the extreme ray representation (or vice versa).  On the other hand, B\"odi et. al \cite{herr09arxiv,bodyhighsymmlpalgo} study how subgroups of RSG consisting of permutation matrices can be utilized to reduce the dimension of linear and integer programs.  The results of B\"{o}di, and other papers explaining how to exploit constraint symmetries to reduce the dimensions of linear programs however, are easily extended the more general notions of symmetries, such as affine and restricted affine symmetries, considered in this manuscript.

In a nutshell, the symmetry exploiting complexity reductions applied to CHM to create symCHM, and the associated parts of the pseudocode in Alg. \ref{symchm} that we be described in the remaining parts of this section in detail are as follows.  As does CHM, symCHM builds a series of increasingly larger inner bounds to the projected polytope by solving carefully selected linear programs over the parent polytope.  However, each step of the inner bounds created by symCHM are forced to include the known symmetries of the projected polyhedron.  This enables memory savings (\S \ref{sec:memred}), by only storing one representative for each equivalence class of symmetric inequalities and extreme rays created by this known symmetry group.  It also enables substantial computational savings (\S \ref{sec:symUp}), as it enables the number of linear programs that must be solved to be reduced from the number of extreme points of the projected polyhedron to the number of equivalence classes (under the known symmetry group) of extreme points.   A second type of complexity reduction (\S \ref{sec:symDD}) is achieved by adapting techniques from Bremner et. al. \cite{Bremner09polyhedralrepresentation} to reduce the complexity of the double descriptions step that must be performed when each new equivalence class (under symmetry) of extreme points are revealed.  Finally, a third type of complexity reduction (\S \ref{sec:symLp}) is achieved, under a method similar to that exploited in B\"{o}di et. al. \cite{herr09arxiv,bodyhighsymmlpalgo}, by reducing the dimension of the linear program necessary to check whether a given equivalence class of facets (under the action of the symmetry group), is terminal.  Exploring further implications of this final idea, in \S \ref{sec:symLpOth}, we point out that in certain network information theory problems focussed not on entire rate regions, but scalar quantities, this final complexity reduction via dimension reduction is especially helpful.

\begin{algorithm}[h]
\caption{Symmetry exploiting CHM}\label{symchm}
\DontPrintSemicolon 
\KwIn{Polyhedron $\mathcal{P}\subseteq \mathbb R^n$, projection dimension $d<n$  and symmetry group $G\leq S_k$ of $\textsf{proj}_k(\mathcal P)$}
\KwOut{Transversal pair $(\mathcal T_{ V},\mathcal T_{ H})$ of $\textsf{proj}_d(\mathcal{P})$}
$(\mathcal T_V,\mathcal T_H)\leftarrow$ \textsf{syminitialhull}($\mathcal{P},k$)\;
\While{$\exists \{\mathbf h\mathbf x\geq \mathbf b\}\in \mathcal T_H$  s.t. \textsf{isterminal}($\mathcal{P},\mathbf{h},\mathbf{b})= 0$ \label{w2_1}} {
$\mathbf{v}\leftarrow$ \textsf{extremepoint}($\mathcal{P},\mathbf{h}$)\;\label{line3}
$\mathcal T_H\leftarrow$\textsf{symupdatehull}($\mathbf{v},V,H$)\;
$\mathcal T_V\leftarrow \mathcal T_V\cup\{\mathbf{v}\}$\label{w2_2}
}
\Return{$(\mathcal T_V,\mathcal T_H)$}\;
\end{algorithm}

\begin{procedure}[h]
\DontPrintSemicolon 
\KwIn{Transversal pair $(\mathcal T_V,\mathcal T_H)$, vertex $\mathbf v$ and group $G$}
\KwOut{Transversal $\mathcal T_H'$ associated with $\mathcal T_V'=\mathcal T_V \cup\{\mathbf{v}\}$}
$(\mathcal T_{V_{\mathcal C}},\mathcal T_{H_{\mathcal C}})\leftarrow \textsf{homog}(\mathcal T_H)^\circ$\;
$\{\mathbf{r}_1,...,\mathbf{r}_t\}\leftarrow$ symDD($\mathcal T_{V_{\mathcal C}},\mathcal T_{H_{\mathcal C}},\{(1 \hspace{2mm} \mathbf v^T)\mathbf{y}\leq  0\},G$)\;
$\mathcal T_H'\leftarrow\{\textsf{ineq}(\mathbf r_1),\hdots,\textsf{ineq}(\mathbf r_t)\}$\;
\Return{$\mathcal T_H'$}\;
\label{proc:symupdatehull}
\caption{symupdatehull($\mathbf{v},V,H$)}
\end{procedure}

\begin{procedure}[h]
\label{symdd}
\DontPrintSemicolon 
\KwIn{Transversal pair $\mathcal T_{V_{\mathcal C}}, \mathcal T_{H_{\mathcal C}}$ of the set of extreme rays and facets of a cone $\mathcal C\subseteq \mathbb R^{d+1}$, inequality $\mathbf a^T\mathbf x\leq  0$, group $G$}
\KwOut{Transversal of set of extreme rays of $\mathcal C\bigcap\cap_{g\in G}\{\mathbf a^T\mathbf x\leq 0\}^g$}
$V_{\mathcal C}\leftarrow \mathcal T_{V_{\mathcal C}}^G$, $H_{\mathcal C}\leftarrow \mathcal T_{H_{\mathcal C}}^G$\;\label{ln:ddofc1}
$(P,N,Z)\leftarrow \textsf{DD}({V_{\mathcal C}}, {H_{\mathcal C}},\mathbf a)$\;\label{ln:dd}
$V_{\mathcal C_{\mathbf v^{\leq}}}\leftarrow P\cup Z$\;
${H_{\mathcal C_{\mathbf v^{\leq}}}}\leftarrow H_{\mathcal C}\cup \{\mathbf a^T\mathbf x\geq  0\}$\;\label{ln:cineq}
$(V_{\mathcal C_{\mathbf v^=}},H_{\mathcal C_{\mathbf v^=}})\leftarrow \textsf{tightenfacet}(P\cup Z,{H_{\mathcal C_{\mathbf v^{\leq}}}})$\;\label{ln:tighten}
$\mathcal A\leftarrow\textsf{repDD}((V_{\mathcal C_{\mathbf v^=}},\mathbf a,G)$\;\label{ln:repdd}
\For{$\mathbf a'\in \mathcal A$}
{$(P,N,Z)\leftarrow \textsf{DD}(V_{\mathcal C_{\mathbf v^=}},H_{\mathcal C_{\mathbf v^=}},\mathbf a')$\;\label{ln:smalldd}
$V_{\mathcal C_{\mathbf v^=}}\leftarrow P\cup Z$\;
$H_{\mathcal C_{\mathbf v^=}}\leftarrow H_{\mathcal C_{\mathbf v^=}}\cup\{\mathbf a'^T\mathbf x\leq 0\}$
}
$\mathcal T_{V_{\mathcal C'}}\leftarrow \textsf{nonisomorphic}((P\setminus{N^G}) \cup \textsf{lift}(V_{\mathcal C_{\mathbf v^=}},\mathbf a),G)$\;\label{ln:newtrans}
\Return{$\mathcal T_{V_{\mathcal C'}}$}
\caption{symDD($\mathcal T_{V_{\mathcal C}},\mathcal T_{H_{\mathcal C} },\mathbf a^T\mathbf x\leq  \mathbf 0,G$)}
\end{procedure}

\subsection{Space Reduction via Symmetry}\label{sec:memred}
Let $ V$ and $ H$ be the set of vertices and facets respectively of $\textsf{proj}_k(\mathcal P)$. For an affine symmetry $g\in G$ and a vertex $\mathbf v$ of $\textsf{proj}_k(\mathcal P)$ denote by $\mathbf v^g$ to be the vertex to which $\mathbf v$ maps to under action of $g$ and let $\mathbf v^G$, the \emph{orbit} of $\mathbf v$ under $G$, be the set of all vertices to which $\mathbf v$ can map to under action of $G$.  $\mathbf v^G$ contains all vertices that are $G$-\textit{equivalent} to $\mathbf v$. The set of all orbits in $V$ under action of $G$ forms a partition of $ V$ and is denoted as $\mathcal O_V$.  Since each facet is simply the convex hull of a set of vertices, the action of the ASG can be extended to $H$ i.e. we define the orbit of a facet $h\in H$ denoted as $h^G$ and $\mathcal O_H$ to be set of all orbits of facets.  The transversal $\mathcal T$ of a set of orbits $\mathcal O$ is a set containing one representative per orbit in $\mathcal O$, and transversals of $\mathcal O_V$ of $\mathcal O_H$ are denoted as $\mathcal T_V$ and $\mathcal T_H$ respectively. Based on the aforementioned McMullen's Upper Bound, it is possible that the size of the double description of $\textsf{proj}_k(\mathcal P)$ is prohibitively large. In this case, we can trade space requirement for orbit computation, which is a basic procedure in computational group theory \cite{seressbook}, by storing only the transversals of the inequality and vertex orbit sets.

\subsection{Exploiting Symmetry to Solve Fewer Linear Programs}\label{sec:symUp}
The transversal $\mathcal T_H$ of the facets of the inner bound carries with it an indicator variable indicating if it is terminal or non-terminal.  At an intermediate step in the algorithm, a non-terminal facet $\mathbf{c}^T \mathbf{x} \geq b$ is selected from the current inner bound's facet transversal $\mathcal T_H$.  Just  as in CHM, the linear program $\min_{\mathbf{y} \in \mathcal P} [\mathbf{c}^T, \mathbf{0}_{d-k}^T] \mathbf{y}$ is solved, and if the result is $b$, the facet is marked as terminal.  If the result is not $b$, the projection of the extreme point attaining the minimum, $\textrm{proj}_k \arg \min_{\mathbf{y} \in \mathcal P} [\mathbf{c}^T, \mathbf{0}_{d-k}^T] \mathbf{y}$, is added to the transversal $\mathcal T_V$.  This act of adding this single extreme point $\mathbf{v}$ to the inner bound's vertex transversal has the same effect as having added the entire orbit $\mathbf{v}^G$ to the full list of extreme points in CHM.  In ordinary CHM, to find these extreme points, $|\mathbf{v}^G|$ linear programs would have had to be solved, whereas in symCHM, only one LP is required to obtain all of them, followed by an orbit computation.
Similarly, if a facet of an inner bound is found to be terminal, all the facets is its orbit under $G$ can be labeled as terminal, amounting to further reduction in the number of LPs solved.

\begin{figure}
\begin{center}
\includegraphics[scale=1.4]{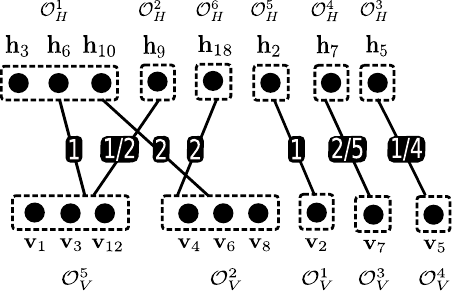}
\end{center}
\caption{Complement of the incidence graph between vertex orbits and facet orbits of the symmetric initial inner bound $\mathcal B_1^G$ obtained by $\textsf{syminitialhull}(\cdot)$ procedure.}\label{fig:syminithull}
\end{figure}

\exmp{[Fewer LPs for Ex. \ref{ex:run} ] 
In case of the example in Fig. \ref{fig:idsc_ex}, as stated in Ex. \ref{ex:netSymGroup}, the network symmetry group $G$ is of order $6$, generated by permutations $(4,5)$ and $(5,6)$. To begin with, the knowledge of symmetry can be used to make the initial inner bound symmetric. Thus all permutations of vertices of initial inner bound $\mathcal B_1$ under the network symmetry group an be added to $\mathcal B_1$ and $\mathcal B_1$ is augmented each time using an iteration of the double description method much on the lines of the inner bound updates in CHM. Tables \ref{tab:syminithull_v} and \ref{tab:syminithull_h} show the vertex and facet orbits of the symmetric initial inner bound $\mathcal B_1^G$ so constructed.  The key complexity reduction here is that additional vertices are obtained via simple orbit computation as opposed to substantially more expensive linear programs. Fig. \ref{fig:syminithull} shows vertex facet incidences between vertex orbits and facet orbits of $\mathcal B_1^G$.
}

\begin{table}[h]
\vspace{3mm}
\begin{center}
\begin{tabular}{|c|l|}
\hline
Orbit Label & Member vertices\\
\hline\hline
$\mathcal O_V^1 $& $\begin{aligned}\mathbf v_2&=\left(0,0,0,0,0,0\right)\end{aligned}$   \\
\hline
$\mathcal O_V^2$ & $\begin{aligned}\mathbf v_4&=(0,0,0,0,1,0)\\\mathbf v_6&=(0,0,0,1,0,0)\\\mathbf v_{8} &= (0,0,0,0,0,1)\end{aligned} $  \\
\hline
$\mathcal O_V^3 $& $\begin{aligned}\mathbf v_7&=\left(0,0,\frac{2}{5},\frac{1}{5},\frac{1}{5},\frac{1}{5}\right)\end{aligned}$   \\
\hline
$\mathcal O_V^4 $& $\begin{aligned}\mathbf v_5&=\left(0,\frac{1}{4},0,\frac{1}{4},\frac{1}{4},\frac{1}{4}\right)\end{aligned}$   \\
\hline
$\mathcal O_V^5$ & $\begin{aligned}\mathbf v_1&=\left(\frac{1}{2},0,0,0,\frac{1}{2},0\right)\\\mathbf v_3&=\left(\frac{1}{2},0,0,0,0,\frac{1}{2}\right)\\\mathbf v_{12} &= \left(\frac{1}{2},0,0,\frac{1}{2},0,0\right)\end{aligned}$   \\
\hline
\end{tabular}
\vspace{3mm}
\caption{Vertex orbits of the symmetric initial hull $\mathcal B_1^G$ obtained by procedure \textsf{syminitialhull}.}\label{tab:syminithull_v}
\end{center}
\end{table}

\begin{table}[h]
\vspace{3mm}
\begin{center}
\begin{tabular}{|c|r|}
\hline
Orbit Label & Member facets\\
\hline\hline
$\mathcal O_H^1$ & $\begin{aligned}2\omega_2+\omega_3 &\leq 2R_6 & (\mathbf h_3)\\ 2\omega_2+\omega_3 &\leq 2R_4 & (\mathbf h_6)\\2\omega_2+\omega_3 &\leq 2R_5 & (\mathbf h_{10})\end{aligned} $  \\
\hline
$\mathcal O_H^2 $& $\begin{aligned}-\omega_1 &\leq 0 & (\mathbf h_9)\end{aligned}$   \\
\hline
$\mathcal O_H^3 $& $\begin{aligned}-\omega_2 &\leq 0 & (\mathbf h_5)\end{aligned}$   \\
\hline
$\mathcal O_H^4 $& $\begin{aligned}-\omega_3 &\leq 0 & (\mathbf h_7)\end{aligned}$   \\
\hline
$\mathcal O_H^5$ & $\begin{aligned}\omega_1+\omega_2+\omega_3+R_4+R_5+R_6 &\leq 1 & (\mathbf h_2)\end{aligned}$   \\
\hline
$\mathcal O_H^6 $& $\begin{aligned}2\omega_1+6\omega_2+3\omega_3 &\leq 2R_4+ 2R_5+2R_6 & (\mathbf h_{18})\\ \end{aligned}$   \\
\hline

\end{tabular}
\vspace{3mm}
\caption{Facet orbits of the symmetric initial hull $\mathcal B_1^G$ obtained by procedure \textsf{syminitialhull}.}\label{tab:syminithull_h}
\end{center}
\end{table}

While the previous example provided some detail regarding the mechanics of this complexity reduction with the running example, of additional interest is the types of complexity reductions of this type that can be achieved in other problems of interest.  The next example addresses this question.

\begin{figure}
\centering
\includegraphics[width=3in]{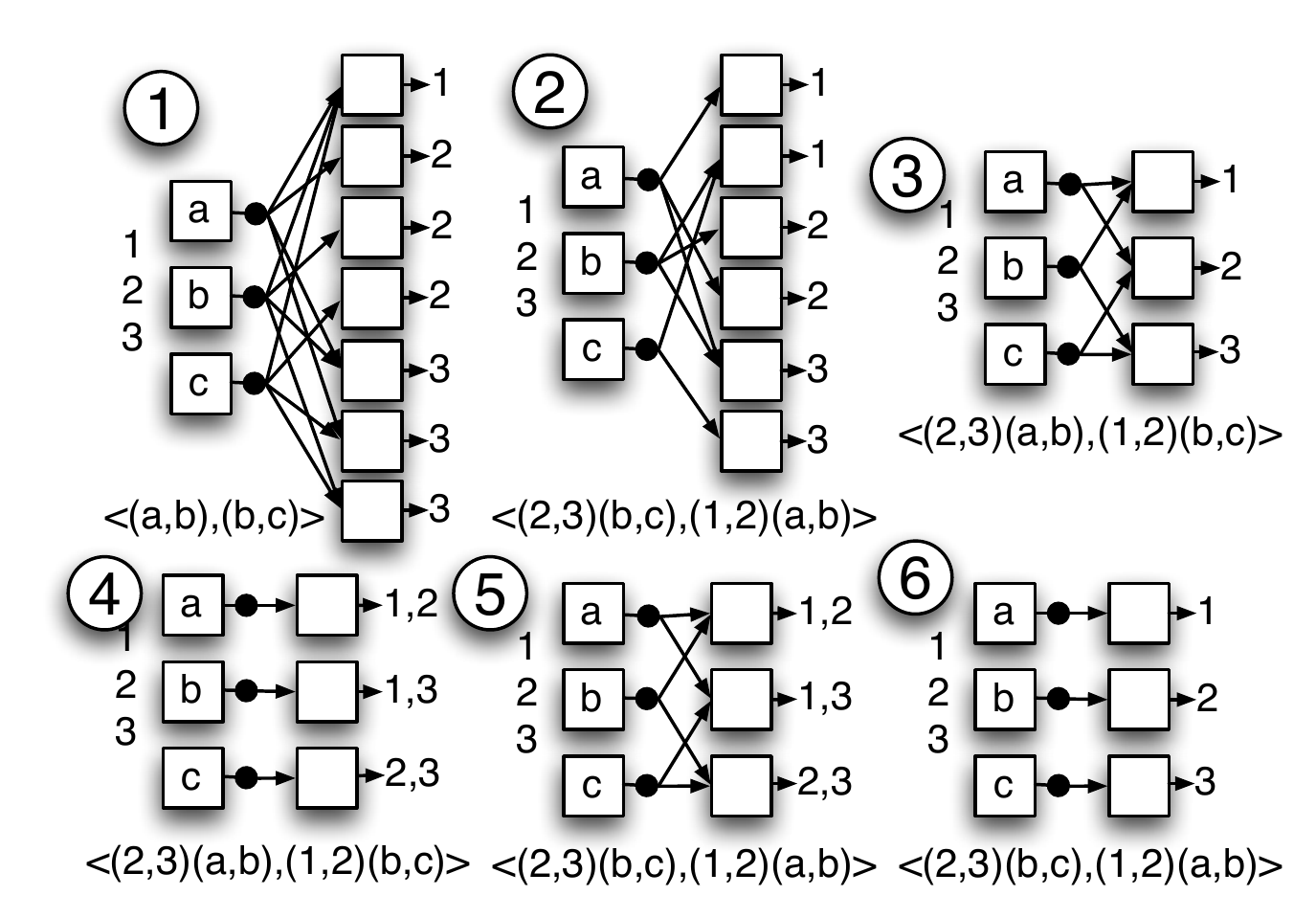}
\caption{Six hypergraph network coding problems that are independent distributed source coding problems that have network symmetry groups of order 6.}\label{fig:sixIDSC}
\end{figure}

\begin{figure}
\centering
\includegraphics[width=3.1in]{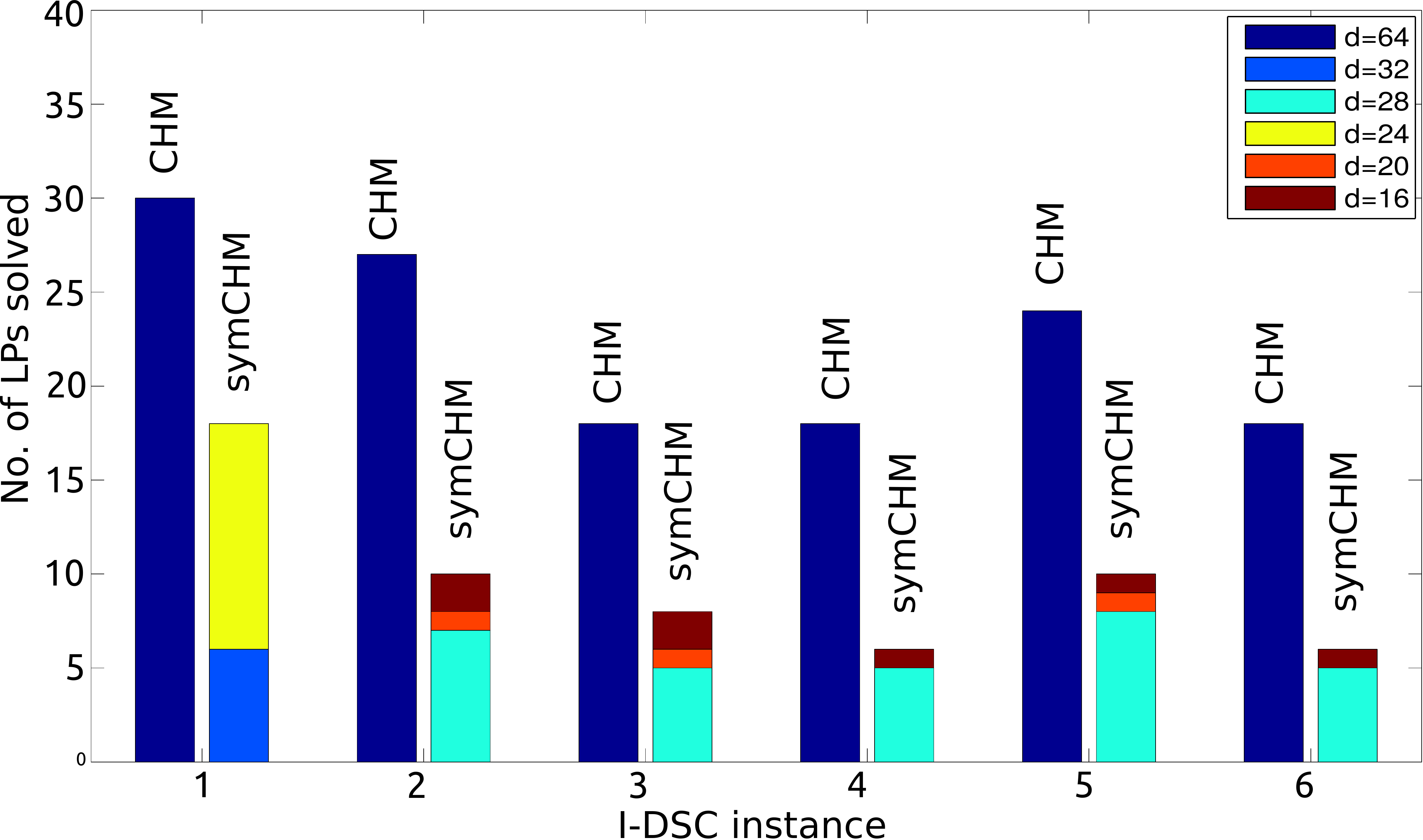}
\caption{The symmetry exploiting complexity reductions described in \S \ref{sec:symUp} and \S \ref{sec:symLp} enable the number of linear programs solved by CHM (left) to be reduced to the number of linear programs solved by symCHM (right).  Furthermore, for checking whether a facet is terminal, the linear program can be reduced to the dimensions indicated by the colors in the legend.}\label{fig:symCHMcr}
\end{figure}

\exmp{[Reduction in $\#$ of LPs] Consider the 6 problems displayed in Fig. \ref{fig:sixIDSC}, which include the running example and 5 other small network coding/IDSC problems, each with network symmetry group of order 6 as displayed under their diagram.  Fig. \ref{fig:symCHMcr} displays the number of linear programs that CHM must solve versus the number of linear programs that symCHM must solve to compute the rate region for these problems.  In each instance, a large reduction is achieved in the number of linear programs that must be solved.}

\subsection{Exploiting Symmetry when Updating the Inner Bound's Double Description Pair}\label{sec:symDD}
 When a new extreme point $\mathbf{v}$ is added to the transversal $\mathcal T_V$, the transversal of the inequalities $\mathcal T_H$ must be updated (proc. \textsf{symupdatehull}) to reflect the new inequalities that the addition of the extreme points $\mathbf{v}^G$ to the symmetric inner bound creates (we call this new polytope the \textit{symmetric improvement}).  In ordinary CHM, this would have been done through of $|\mathbf{v}^G|$ steps of the DD method applied to the complete inequality description of the symmetric inner bound.  However, based on Lemma \ref{chm_incid}, which is the same insight from which the incidence decomposition method \cite{Bremner09polyhedralrepresentation} for representation conversion of symmetric polyhedra is derived, we can perform DD steps of smaller size (proc. \textsf{symDD}) to obtain the new facets that must be added to the transversal. The size of a DD step, in this context is the number of extreme rays in the input double description step.

\leml{Let $\mathcal P_k^{(\ell)}$ be an inner bound on $\textsf{proj}_k(\mathcal P)$ whose ASG has $G_p$ as a subgroup, and let $\mathbf v$ be a new vertex of a symmetric improvement  $\mathcal P_k^{(\ell + 1)}$. If, in $\mathcal P_k^{(\ell + 1)}$, $\{f_1,\hdots,f_t\}$ is the set of facets incident to $\mathbf v$ then,  $\{f_1^g,\hdots,f_t^g\}$ is the set of facets incident to $\mathbf v^g$.}{chm_incid}

\noindent Lemma \ref{chm_incid} ensures that as long as we calculate the facets of $\mathcal P_k^{(\ell+1)}$ incident to $\mathbf v$ correctly, and include any of these facets that are $G$-inequivalent into the new transversal $\mathcal T_H$ after removing those non-terminal inequalities that the new extreme points violate, the new facet transversal will reflect all of the $G$-inequivalent facets of $\mathcal P_k^{(\ell +1)}$.  The key issue in calculating the facets incident to $\mathbf v$ in $\mathcal P_k^{(\ell +1)}$ correctly is that there may be some vertices in $\mathbf v^G \setminus \{\mathbf v\}$ that are adjacent to $\mathbf v$.

To address this issue, we first determine the double description pair of cone $\mathcal C_{\mathbf v^=}$, given as,
\begin{equation}
\mathcal{C}_{\mathbf{v}^=} \triangleq \mathcal{C} \cap \{ [1\ \mathbf{v}^T] \mathbf{x} = 0\} ,
\end{equation}
where $\mathcal C$ is the homogenized polar of the current inner bound $\mathcal{C} = \textrm{homog}(\mathcal P_k^{\ell})^{\circ}$. The double description pair associated with $\mathcal C$ is computed from the associated transversals in line \ref{ln:ddofc1} of proc. \textsf{symDD}. The double description pair of $\mathcal{C}_{\mathbf{v}^=}$ is determined by first determining the double description pair of $\mathcal C_{\mathbf v^\leq}=\mathcal C\cap \{ (1, \mathbf{v}^T) \mathbf{x} \leq 0\}$  through an iteration of double description step, in lines \ref{ln:dd} and \ref{ln:cineq} of proc. \textsf{symDD}. The procedure $\textsf{DD}$, used for this purpose, returns sets $P,Z$ and $N$ of extreme rays, where $P$ and $N$ are the extreme rays of $\mathcal C$ that strictly satisfy and violate the inequality $\{ (1, \mathbf{v}^T) \mathbf{x} \leq 0\}$ respectively, while the set $Z$ contains the extreme rays of $\mathcal C$ that evaluate to zero w.r.t. $\{ (1\ \mathbf{v}^T) \mathbf{x} \leq 0\}$ in addition to the new extreme rays that are computed as conic combinations of rays in $P$ and $N$.  The set of extreme rays of $\mathcal C_{\mathbf v^=}$ is the set $Z$, i.e. the rays of $\mathcal C_{\mathbf v\geq}$ that have inequality $\{ (1, \mathbf{v}^T) \mathbf{x} = 0\}$ incident to them, while its inequality representation is formed by the subset of inequalities in $\hat{H_{\mathcal C}}$ that are adjacent to $\{ (1, \mathbf{v}^T) \mathbf{x} \leq 0\}$. The computation of double description pair of $\mathcal C_{\mathbf v^=}$ is condensed into proc. \textsf{tightenfacet} on line \ref{ln:tighten} of proc. \textsf{symDD}. Note that cone $\mathcal C_{\mathbf v^=}$ has dimension one lower that of $\mathcal C_{\mathbf v^\leq}$ along with having only a subset of its extrem rays and facets. The symmetry exploitation is achieved by using only  $\mathcal C_{\mathbf v^=}$ from this poin onwards, to compute the symmetric update. To check to see if any vertices in $\mathbf v^G\setminus \{\mathbf v\}$ are adjacent to $\mathbf v$, and if so, which ones, we can determine the set $\mathcal A$, defined as,
\begin{equation}\label{eq:Aset}
\mathcal{A} = \left\{ (1,\mathbf z^T) \left| \mathbf{z} \in \mathbf{v}^G\setminus \{\mathbf{v}\}  \min_{\mathbf{x} \in \mathcal{C}_{\mathbf{v}^=}} (1\ \mathbf{z}^T) \mathbf{x} < 0 \right. \right\}
\end{equation}
Procedure \textsf{repDD} (line \ref{ln:repdd} of \textsf{symDD}) is used to compute the set $\mathcal A$.
The rays of $V_{\mathcal C_{\mathbf v^=}}$ are further refined by adding the inequalities $\{ \mathbf{w}^T \mathbf{x} \leq 0\}$ for each $\mathbf{w} \in \mathcal{A}$ if any, through $|\mathcal{A}|$ further DD steps. This refinement is carried out in line \ref{ln:smalldd} of proc. \textsf{symDD}. The new inequality transversal of $\mathcal{P}_{k}^{(\ell+1)}$ is can be created by removing any $G$-equivalent inequalities from $\mathcal{P}_k^{\ell}$'s inequality description (i.e. rays in the homogenized polar) in these $| \mathcal{A} | + 1$ DD steps, and by adding the representatives of the new rays introduced at the end of these $|\mathcal{A}|+1$ DD steps, in line \ref{ln:newtrans} of proc. \textsf{symDD}, where proc. \textsf{lift} is used to embed the refined set of rays of $\mathcal C_{\mathbf v^=}$ into the higher dimensional space in which $\mathcal C$ existed.
 The consideration of symmetry in this step of updating the inequality description of the inner bound, reduced both the number of DD steps required for CHM and their size.  Indeed, only $|\mathcal{A}|+1$ DD steps must be performed to find the result of adding $|\mathbf{v}^G|$ new extreme points.  Also, the size of each these DD steps is substantially smaller, since the cone $\mathcal{C}_{\mathbf{v}^{=}}$ is being dealt with in the $|\mathcal{A}|$ latter DD steps, rather than $\mathcal{C}$.

\begin{figure*}
\begin{center}
\includegraphics[scale = 0.5]{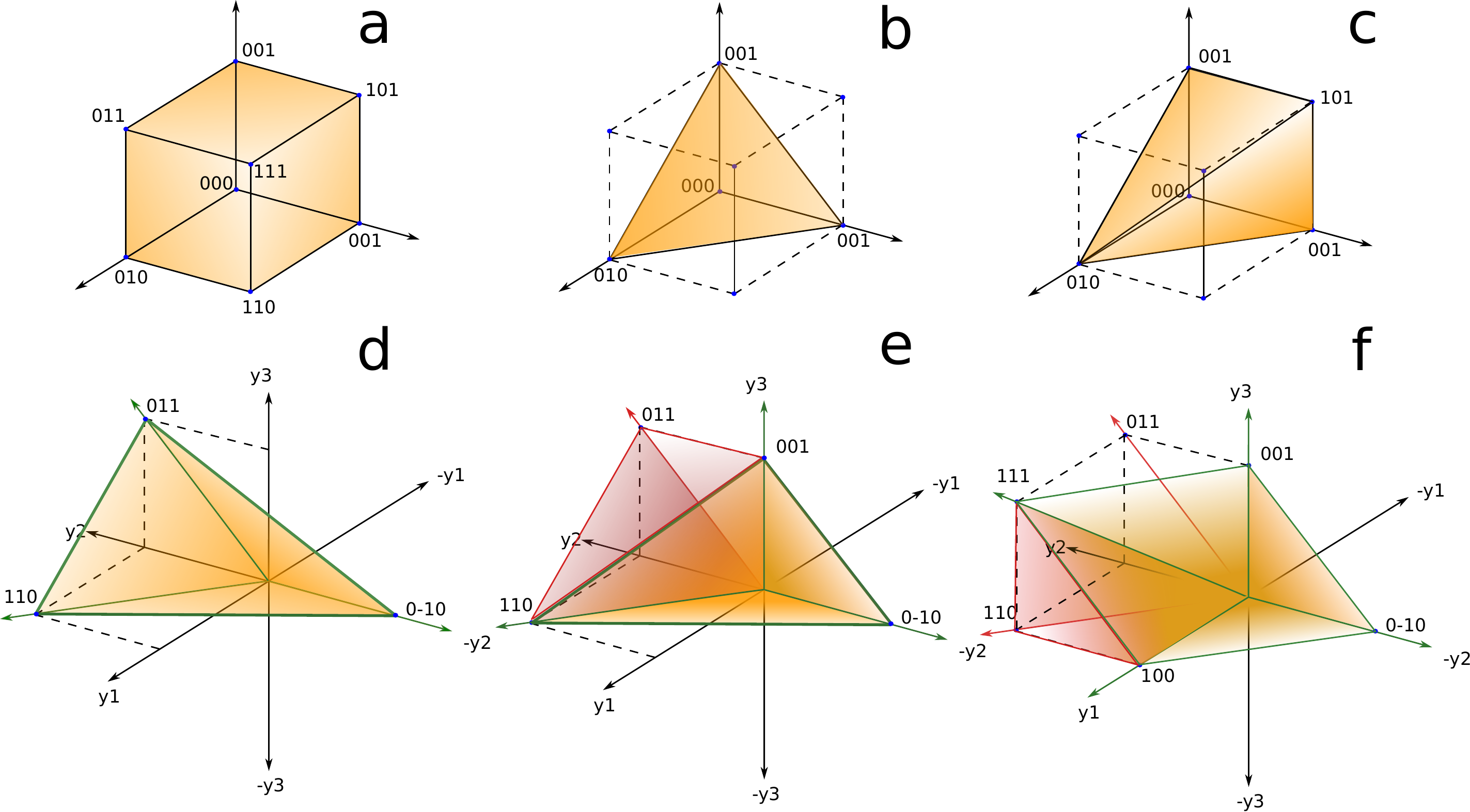}
\vspace{2mm}
\caption{Symmetric update of an inner bound of a 3D cube. Part (a) shows the 3D cube in question with symmetry group $S_3$ which is the projection polytope to be computed, part (b) shows a 3D simplex that forms a symmetric inner bound to the projection, by the virtue of a symmetry group $S_3$. The inequalities and extreme points of (c) are shown in \eqref{eq:simplexv} and \eqref{eq:simplexh} respectively. Part (c) shows the updated inner bound obtained by adding vertex $\mathbf v=(1,0,1)$ to the description. Polar of homogenization of the polytope in (c) lies in $(\mathbb R^4)^\circ$, which is referred to as $\mathcal C_{\mathbf v^{\leq}}$, whose rays are shown in \eqref{eq:cvgeqrays}. The cone in part (d) is $\mathcal C_{\mathbf v^=}$, that lives in $(\mathbb R^3)^\circ$, whose rays are shown in \eqref{eq:cveqrays}.  Parts (e) and (f) show the lower dimentional double description steps performed to obtain the symmetric update. The extreme rays of the cone in part (f) are shown in \eqref{eq:lastconev}}\label{cubeupdate}
\end{center}
\end{figure*}

\exmp{[Symmetric DD Update for a Cube]
This is a simple example elaborating how procedure \textsf{symDD} works. Consider a cube $\mathcal P_3\subseteq \mathbb R^3$ (Fig. \ref{cubeupdate}-a) with inequality representation,
\begin{equation}
H = \{0\leq\mathbf x_i\leq 1,i\in \{1,2,3\}\},
\end{equation}
and extreme points,
\begin{equation}
V=\left\{\begin{aligned} & (0,0,0),(0,0,1),(0,1,0),(0,1,1),\\ & (1,0,0),(1,0,1),(1,1,0),(1,1,1) \end{aligned}\right\}
\end{equation}
$\mathcal P_3$ is the projection of any hypercube $\mathcal P\subseteq \mathbb R^d,d\geq 4$. The symmetric group $S_3$ is a group of symmetries of $\mathcal P_3$, as $\mathcal P_3$ is stabilized setwise under any permutation of coordinate dimensions. A simplex $P_3^{(l)}\subseteq \mathcal P_3\subseteq \mathbb R^3$ (Fig. \ref{cubeupdate}-b) is the convex hull of $V^{(l)}\subseteq V$,
\begin{equation}\label{eq:simplexv}
V^{(l)}=\{(0,0,0),(0,0,1),(0,1,0),(1,0,0)\},
\end{equation}
and has inequality representation,
\begin{equation}\label{eq:simplexh}
H^{(l)} = \{0\leq\mathbf x_i,i\in \{1,2,3\}\}\cup\{\mathbf x_1+\mathbf x_2+\mathbf x_3\leq 1\}.
\end{equation}
Let $\mathcal P_3^{(l)}$ be the inner bound at the $l$th iteration of symCHM. Now consider the problem of computing the inequality representation $H^{(l+1)}$ of the symmetric update $P_3^{(l+1)}$, i.e. the convex hull of $V^{(l)}\cup \mathbf v^{S_3}$ where $\mathbf v$ is the vertex $(1,0,1)$, which is presumably found by solving a linear program over the aforementioned hypercube $\mathcal P$ (line \ref{line3} of symCHM). The first input to procedure $\textsf{symDD}$ is the transversal of the orbits of the rays of $\mathcal C=\textsf{homog}(P_3^{(l)})^\circ\subseteq (\mathbb R^4)^\circ$, given as,
\begin{equation}
\mathcal T_{\mathcal C}=\{ (0,-1,0,0),(-1,1,1,1)\},
\end{equation}
which bears one to one correspondance with transversal of orbits of $S_3$ in $H^{(l)}$. The second input is the inequality $ \mathbf y_0+\mathbf y_1+\mathbf y_3\leq 0$, which corresponds to vertex $\mathbf v$, after homogenization and polar. The initial double description step inserts this inequality in $\mathcal C$, giving the double description pair of cone $\mathcal C\cap\{\mathbf y|\mathbf y_0+\mathbf y_1+\mathbf y_3\leq 0\}\subseteq (\mathbb R^4)^\circ$ which corresponds to the polytope shown in Fig. \ref{cubeupdate}-c, in the original space. $\mathcal C\cap\{\mathbf y|\mathbf y_0+\mathbf y_1+\mathbf y_3\leq 0\}$ has inequality representation,
\begin{equation}
\begin{aligned}
H^{\mathcal C_{\mathbf v\geq}} &= T_1\cup T_2\cup T_3 \text{ where, }\\
 & \begin{aligned} T_1 &=\left\{\mathbf y_0+\mathbf y_i\leq 0,\forall i\in \{1,2,3\}\right\}\\ T_2 &=  \left\{\mathbf y_0\leq 0\right\}\\ T_3 &=  \left\{\sum_{i\in \{0,1,3\}}\mathbf y_i\leq 0\right\}\end{aligned}
\end{aligned},
\end{equation}
where the last inequality is the newly added inequality, and extreme rays,
\begin{equation}\label{eq:cvgeqrays}
V^{\mathcal C_{\mathbf v^{\leq}}}=\left\{\begin{aligned}& (0,-1,0,0),(0,0,0,-1),(0,0,-1,0),\\
& (-1,0,1,1),(-1,1,1,0)\end{aligned}\right\}
\end{equation}
The first two rays in $V^{\mathcal C_{\mathbf v^{\leq}}}$ belong to the set $P$ in line \ref{ln:dd}  of \textsf{symDD} and last three rays belong to set $Z$, while the set $N$ contains the ray $(-1,1,1,1)$, which in original space corresponds to the inequality $\mathbf x_1+\mathbf x_2+\mathbf x_3\leq 1$, is not part of the inequality representation of the symmetric updates, and is to be removed from consideration along with all its permutations in line \ref{ln:newtrans} at the end of proc. \textsf{symDD}($N^{S_3}=N$ in this case).
The cone $\mathcal C_{\mathbf v^{=}}$ is obtained by making last inequality in $H^{\mathcal C}$ tight, and has inequality representation,
\begin{equation}
H^{\mathcal C_{\mathbf v^{=}}}=\left\{\begin{aligned} & -\mathbf y_1\leq 0,-\mathbf y_1+\mathbf y_2-\mathbf y_3\leq 0, \\
& -\mathbf y_3\leq 0,-\mathbf y_1-\mathbf y_2\leq 0\end{aligned}\right\},
\end{equation}
which is obtained by substituting $\mathbf y_0=-\mathbf y_1-\mathbf y_3$ in the first 4 inequalities of $H^{\mathcal C}$. Note that the last inequality in $H^{\mathcal C_{\mathbf v^{=}}}$ is redundant, i.e. the cone $\mathcal C_{\mathbf v^{=}}$ remains unchanged even after removing this inequality. Such redundancy can be detected, using e.g. the algebraic adjacency oracle described by Fukuda et al. in \cite{fukuda1996double}. In the framework of Bremner et. al.\cite{Bremner09polyhedralrepresentation}, this would require solving of a linear program, a situation which is alleviated in our case by the knowledge of both descriptions of $\mathcal C$.
Cone $\mathcal C_{\mathbf v^{=}}$ is shown in Fig. \ref{cubeupdate}-d, and has extreme rays obtained from set $Z$ in line \ref{ln:dd} of \textsf{symDD},
\begin{equation}\label{eq:cveqrays}
V^{\mathcal C_{\mathbf v^=}}=\{(1,1,0),(0,-1,0),(0,1,1)\}.
\end{equation}
Next, we form inequalities $\mathbf y_0+\mathbf y_2+\mathbf y_3\leq 0$ and $\mathbf y_0+\mathbf y_1+\mathbf y_2\leq 0$, corresponding to other vertices in the orbit of $\mathbf v$ i.e. $\mathbf v^G\setminus\{\mathbf v\}=\{(0,1,1),(1,1,0)\}$.
Substituting $\mathbf y_0=-\mathbf y_1-\mathbf y_3$ in these inequalities, we get inequalities $-\mathbf y_1+\mathbf y_2\leq 0$ and $\mathbf y_2-\mathbf y_3\leq 0$. The set $\mathcal A$ is formed by checking if any rays in $V^{\mathcal C_{\mathbf v^=}}$ fail to satisfy these inequalities. In this case, ray $(0,1,1)$ fails inequality $-\mathbf y_1+\mathbf y_2\leq 0$ and ray $(1,1,0)$ fails inequality $\mathbf y_2-\mathbf y_3\leq 0$, implying that $\mathcal A=\{(1,0,1,1),(1,1,1,0)\}$. Standard DD steps are now performed (line \ref{ln:smalldd} of \textsf{symDD}), for inserting inequalities $-\mathbf y_1+\mathbf y_2\leq 0$ (obtained by substituting $\mathbf y_0=-\mathbf y_1-\mathbf y_3$ in $\mathbf y_0+\mathbf y_2+\mathbf y_3\leq 0$) and $\mathbf y_2-\mathbf y_3\leq 0$ in $\mathcal C_{\mathbf v^{=}}$ (obtained by substituting $\mathbf y_0=-\mathbf y_1-\mathbf y_3$ in $\mathbf y_0+\mathbf y_1+\mathbf y_2\leq 0$), as shown in Fig.  \ref{cubeupdate}-e and Fig. \ref{cubeupdate}-f respectively. The cone in Fig. \ref{cubeupdate}-f has extreme rays,
\begin{equation}\label{eq:lastconev}
\{(0,-1,0),(0,0,1),(1,1,1),(1,0,0)\},
\end{equation}
which after prepending coordinate $\mathbf y_0=-\mathbf y_1-\mathbf y_3$ become,
\begin{equation}
\left\{\begin{aligned} & (0,0,-1,0),(-1,0,0,1),\\
& (-2,1,1,1),(-1,1,0,0)\end{aligned}\right\} \subseteq (\mathbb R^4)^\circ
\end{equation}
In the original space, these correspond to the inequalities,
\begin{equation}\label{eq:incidentineq}
\left\{-\mathbf x_2\leq 0,\mathbf x_3\leq 1, \mathbf x_1+\mathbf x_2+\mathbf x_3\leq 2,\mathbf x_1\leq 1\right\}
\end{equation}
which are the inequalities incident to vertex $\mathbf v=(1,0,1)$ in the symmetric update. The inequalities incident to rays in $\mathbf v^G\setminus \{\mathbf v\}$ are all obtained as permutations of those incident to $\mathbf v$, according to lemma \ref{chm_incid}. Finally, the inequality description of the symmetric update is obtained by permuting inequalities in \eqref{eq:incidentineq} under $S_3$ and taking union with the inequalities associated with set $P\setminus N^{S_3}$.
\begin{equation}
H^{(l+1)}=\left\{0\leq \mathbf x_i\leq 1,\forall i\in\{1,2,3\}\right\}\cup\left\{\sum_{i\in\{1,2,3\}}\mathbf x_i\leq 2\right\}
\end{equation}
}

The previous example was included for the purposes of demonstrating the need for consideration of adjacency of different elements of the orbit of the newly discovered extreme point.  The next example continues the demonstration of the running example \ref{ex:run} for the symmetric bound update double descriptions step.

\begin{figure}
\begin{center}
\includegraphics[width=3.15in]{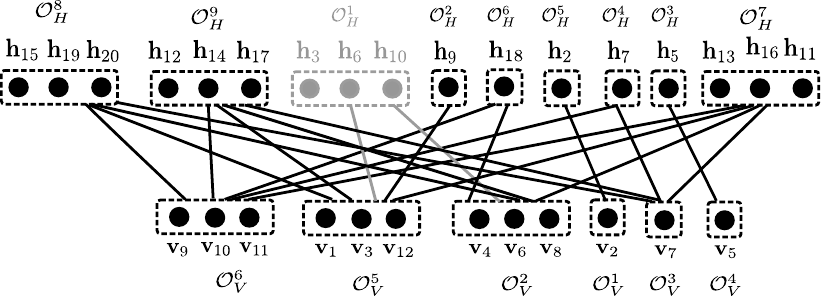}
\end{center}
\caption{Complement of the incidence graph between vertex orbits and facet orbits of the second symmetric inner bound $\mathcal B_2^G$ obtained by $\textsf{symDD}(\cdot)$ procedure. This bound is in fact the final bound, as all its faces are terminal.}\label{fig:symsecond}
\end{figure}

\exmp{[Symmetric Update for Ex. \ref{ex:run}]
the running example in Fig. \ref{fig:idsc_ex}, after construction the initial symmetric inner bound $\mathcal B_1^G$, consider the situation where a new vertex $\mathbf v_{10}$ is obtained by solving a linear program. The orbit of $\mathbf v_{10}$, i.e. $\mathbf v_1^G$ where $G$ is the network symmetry group, contains two additional vertices $\mathbf v_{9}$ and $\mathbf v_{11}$. The problem we now face is that of constructing the inequality description of convex hull of vertices of $\mathcal B_1^G$ and $\mathbf V_{10}^G$.  In the procedure $\textsf{symDD}(\cdot)$, an initial asymmetric double description step is carried out to compute the inequality description of $\mathcal B_1^G$ augmented by $\mathbf v_{10}$. Following equations illustrate the construction of new inequalities that hold for the inner bound obtained by augmenting $\mathcal B_1^G$ by vertex $\mathbf v_{10}$ via an iteration of the double description method,
\begin{equation}
\begin{aligned}
\frac{1}{2}*&(2\omega_2 +& \omega_3 & &-2R_6&\leq 0 )& (\mathbf h_3)\\
+\frac{1}{2}*&(2\omega_2 +& \omega_3 &-2R_4& &\leq 0 )& (\mathbf h_6)\\
 \cline{1-6}
& 2\omega_2 +& \omega_3 &-R_4 &-R_6&\leq 0 & (\mathbf h_{12})
\end{aligned}
\end{equation}
\begin{equation}
\begin{aligned}
\frac{1}{2}*&( & -\omega_3 & &\leq 0 )& (\mathbf h_7)\\
+\frac{1}{2}*&(2\omega_2 +& \omega_3 &-2R_4&\leq 0 )& (\mathbf h_6)\\
 \cline{1-6}
& \omega_2 +&  &-R_4 &\leq 0 & (\mathbf h_{13})
\end{aligned}
\end{equation}
\begin{equation}
\begin{aligned}
\frac{1}{2}*&(2\omega_2 +& \omega_3 & &-2R_5&\leq 0 )& (\mathbf h_{10})\\
+\frac{1}{2}*&(2\omega_2 +& \omega_3 &-2R_4& &\leq 0 )& (\mathbf h_6)\\
 \cline{1-6}
& 2\omega_2 +& \omega_3 &-R_4 &-R_5&\leq 0 & (\mathbf h_{14})
\end{aligned}
\end{equation}
\begin{equation}
\begin{aligned}
\frac{1}{2}*&(2\omega_1+&6\omega_2 +& 3\omega_3 & -2R_4 &-2R_5 & -2R_6 &\leq 0 )& (\mathbf h_{18})\\
+\frac{1}{2}*&(&2\omega_2 +& \omega_3 &-2R_4& & &\leq 0 )& (\mathbf h_6)\\
 \cline{1-8}
&\omega_1&+ 4\omega_2 +& 2\omega_3 &-2R_4 &-R_5 &-R_6 &\leq 0 & (\mathbf h_{15})
\end{aligned}
\end{equation}

\begin{figure}
\begin{center}
\includegraphics[scale=1.4]{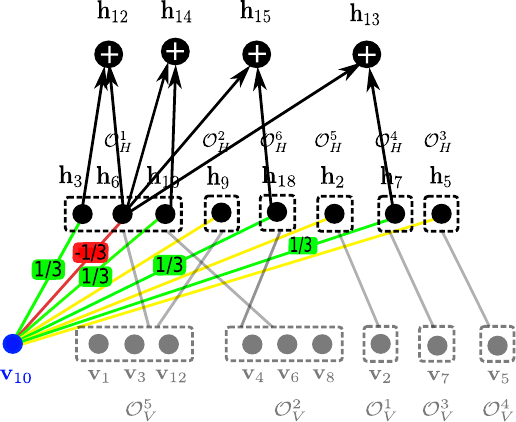}
\end{center}
\caption{Illustration of the asymmetric Double Description step in line \ref{ln:dd} of procedure $\textsf{symDD}(\cdot)$. The colors red, yellow and green represent violation, equality, or strict satisfaction of a particular inequality by $\mathbf v_{10}$. The inequality violated by $\mathbf v_{10}$ i.e. $\mathbf h_6$ is combined with each inequality strictly satisfied by $\mathbf v_{10}$ i.e. $\mathbf h_2,\mathbf h_3,\mathbf h_{10}$, and $\mathbf h _{2}$, to produce the new inequalities that hold for the augmented inner bound i.e. $\mathbf h_{12},\mathbf h_{14},\mathbf h_{15},\mathbf h_{13}$.}\label{symdd_asym}
\end{figure}

Fig. \ref{symdd_asym} graphically describes the asymmetric double description step just carried out. We call the resultant asymmetric inner bound $\mathcal B_2$.  After the asymmetric double description step, the cone $\mathcal C_{\mathbf v_{10}^=}$ is constructed, as described in table \ref{tab:embed_cveq}. Additional double description steps, if needed, are now carried out on the cone $\mathcal C_{\mathbf v_{10}^=}$. In this case, no additional steps are needed as the set $\mathcal A$ (given in \eqref{eq:Aset}) is empty. Thus the new symmetric inner bound $\mathcal B_2^G$ can be constructed by simply considering permutations of facets incident to $\mathbf v_{10}$ in $\mathcal B_2$. The vertex and facet orbits of $\mathcal B_2^G$ so constructed are described in tables \ref{tab:symsecond_v} and \ref{tab:symsecond_h} respectively, while Fig. \ref{fig:symsecond} shows the vertex-facet orbit incidences of $\mathcal B_2^G$.
}

\begin{table*}
\vspace{3mm}
\begin{center}
\begin{tabular}{|l|c|c|}
\hline
Vertex of $\mathcal B_2$ adjacent to $\mathbf v_{10}$ & Homogenized Polar Inequality & $\mathcal C_{\mathbf v_{10}^=}$ Inequality\\
\hline\hline
$\mathbf v_1=\left(\frac{1}{2},0,0,0,\frac{1}{2},0\right)$ & $\omega^\circ+\frac{1}{2}\omega_1^\circ+\frac{1}{2}R_5^\circ\leq 0$ & $\frac{1}{2}\omega_1^\circ-\frac{1}{3}\omega_3^\circ+\frac{1}{6}R_5^\circ-\frac{1}{3}R_6^\circ\leq 0$  \\
\hline
$\mathbf v_3=\left(\frac{1}{2},0,0,0,0,\frac{1}{2}\right)$ & $\omega^\circ+\frac{1}{2}\omega_1^\circ + \frac{1}{2}R_6^\circ\leq 0$ & $\frac{1}{2}\omega_1^\circ -\frac{1}{3}\omega_3^\circ - \frac{1}{3}R_5^\circ + \frac{1}{6}R_6^\circ\leq 0$    \\
\hline
$\mathbf v_5=\left(0,\frac{1}{4},0,\frac{1}{4},\frac{1}{4},\frac{1}{4}\right)$ & $\omega^\circ+\frac{1}{4}\omega_2^\circ + \frac{1}{4}R_4^\circ + \frac{1}{4}R_5^\circ+ \frac{1}{4}R_6^\circ\leq 0$ & $\frac{1}{4}\omega_2^\circ - \frac{1}{3}\omega_3^\circ + \frac{1}{4}R_4^\circ - \frac{1}{12}R_5^\circ - \frac{1}{12}R_6^\circ\leq 0$  \\
\hline
$\mathbf v_4=(0,0,0,0,1,0)$ & $\omega^\circ+R_5^\circ\leq 0$ & $-\frac{1}{3}\omega_3^\circ+\frac{2}{3}R_5^\circ -\frac{1}{3}R_6^\circ\leq 0$   \\
\hline
$\mathbf v_{8} = (0,0,0,0,0,1)$ & $\omega^\circ+R_6^\circ\leq 0$ & $-\frac{1}{3}\omega_3^\circ-\frac{1}{3}R_5^\circ +\frac{2}{3}R_6^\circ\leq 0$   \\
\hline
$\mathbf v_2=\left(0,0,0,0,0,0\right)$ & $\omega^\circ\leq 0$ &  $-\frac{1}{3}\omega_3^\circ-\frac{1}{3}R_5^\circ -\frac{1}{3}R_6^\circ\leq 0$   \\
\hline
$\mathbf v_7=\left(0,0,\frac{2}{5},\frac{1}{5},\frac{1}{5},\frac{1}{5}\right)$ & $\omega^\circ+\frac{2}{5}\omega_3^\circ +\frac{1}{5}R_4^\circ + \frac{1}{5}R_5^\circ+ \frac{1}{5}R_6^\circ\leq 0$ & $\frac{1}{15}\omega_3^\circ +\frac{1}{5}R_4^\circ - \frac{2}{15}R_5^\circ - \frac{2}{15}R_6^\circ\leq 0$   \\
\hline
\end{tabular}
\vspace{3mm}
\caption{Construction of cone $\mathcal C_{\mathbf v_{10}^=}$. A new dummy variable $\omega$ is used to correspond to the extra dimension added by homogenization. $\circ$ on top of a variable corresponds to the polar dual of that variable. Thus, a vertex $\mathbf v_1=\left(\frac{1}{2},0,0,0,\frac{1}{2},0\right)$ is homogenized to get a ray $\left(1,\frac{1}{2},0,0,0,\frac{1}{2},0\right)$ which is interpreted as an inequality in the second column under polar duality. Vertex $\mathbf v_{10}=\left(0,0,\frac{1}{3},0,\frac{1}{3},\frac{1}{3}\right)$ under the same mapping yiends inequality $\omega^\circ + \frac{1}{3}\omega_3^\circ+\frac{1}{3}R_5^\circ +\frac{1}{3}R_6^\circ\leq 0$, which is made tight to give $\omega^\circ = -\frac{1}{3}\omega_3^\circ-\frac{1}{3}R_5^\circ -\frac{1}{3}R_6^\circ$. It is then substituted in column 2 inequalities to yield column 3 inequalities describing $\mathcal C_{\mathbf v_{10^=}}$}\label{tab:embed_cveq}
\end{center}
\end{table*}

\begin{table}[h]
\vspace{3mm}
\begin{center}
\begin{tabular}{|c|l|}
\hline
Orbit Label & Member vertices\\
\hline\hline
$\mathcal O_V^1 $& $\begin{aligned}\mathbf v_2&=\left(0,0,0,0,0,0\right)\end{aligned}$   \\
\hline
$\mathcal O_V^2$ & $\begin{aligned}\mathbf v_4&=(0,0,0,0,1,0)\\\mathbf v_6&=(0,0,0,1,0,0)\\\mathbf v_{8} &= (0,0,0,0,0,1)\end{aligned} $  \\
\hline
$\mathcal O_V^3 $& $\begin{aligned}\mathbf v_7&=\left(0,0,\frac{2}{5},\frac{1}{5},\frac{1}{5},\frac{1}{5}\right)\end{aligned}$   \\
\hline
$\mathcal O_V^4 $& $\begin{aligned}\mathbf v_5&=\left(0,\frac{1}{4},0,\frac{1}{4},\frac{1}{4},\frac{1}{4}\right)\end{aligned}$   \\
\hline
$\mathcal O_V^5$ & $\begin{aligned}\mathbf v_1&=\left(\frac{1}{2},0,0,0,\frac{1}{2},0\right)\\\mathbf v_3&=\left(\frac{1}{2},0,0,0,0,\frac{1}{2}\right)\\\mathbf v_{12} &= \left(\frac{1}{2},0,0,\frac{1}{2},0,0\right)\end{aligned}$   \\
\hline
$\mathcal O_V^6 $& $\begin{aligned}\mathbf v_9&=\left(0,0,\frac{1}{3},\frac{1}{3},\frac{1}{3},0\right)\\ \mathbf v_{10}&=\left(0,0,\frac{1}{3},0,\frac{1}{3},\frac{1}{3}\right)\\\mathbf v_{11}&=\left(0,0,\frac{1}{3},\frac{1}{3},0,\frac{1}{3}\right) \end{aligned}$   \\
\hline
\end{tabular}
\vspace{3mm}
\caption{Vertex orbits of the second symmetric inner bound $\mathcal B_2^G$ obtained by procedure $\textsf{symDD}(\cdot)$.}\label{tab:symsecond_v}
\end{center}
\end{table}

\begin{table}[h]
\vspace{3mm}
\begin{center}
\begin{tabular}{|c|r|}
\hline
Orbit Label & Member facets\\
\hline\hline
$\mathcal O_H^1$ & Deemed non-terminal  \\
\hline
$\mathcal O_H^2 $& $\begin{aligned}-\omega_1 &\leq 0 & (\mathbf h_9)\end{aligned}$   \\
\hline
$\mathcal O_H^3 $& $\begin{aligned}-\omega_2 &\leq 0 & (\mathbf h_5)\end{aligned}$   \\
\hline
$\mathcal O_H^4 $& $\begin{aligned}-\omega_3 &\leq 0 & (\mathbf h_7)\end{aligned}$   \\
\hline
$\mathcal O_H^5$ & $\begin{aligned}\omega_1+\omega_2+\omega_3+R_4+R_5+R_6 &\leq 1 & (\mathbf h_2)\end{aligned}$   \\
\hline
$\mathcal O_H^6 $& $\begin{aligned}2\omega_1+6\omega_2+3\omega_3 &\leq 2R_4+ 2R_5+2R_6 & (\mathbf h_{18})\\ \end{aligned}$   \\
\hline
$\mathcal O_H^7 $& $\begin{aligned}\omega_2 &\leq R_4& (\mathbf h_{13})\\ \omega_2 &\leq R_5& (\mathbf h_{16})\\\omega_2 &\leq R_6& (\mathbf h_{11})\\ \end{aligned}$   \\
\hline
$\mathcal O_H^8 $& $\begin{aligned}\omega_1+4\omega_2+2\omega_3 &\leq 2R_4+ R_5+R_6 & (\mathbf h_{15})\\\omega_1+4\omega_2+2\omega_3 &\leq R_4+ 2R_5+R_6 & (\mathbf h_{19})\\\omega_1+4\omega_2+2\omega_3 &\leq R_4+ R_5+2R_6 & (\mathbf h_{20})\\ \end{aligned}$   \\
\hline
$\mathcal O_H^9 $& $\begin{aligned}2\omega_2+\omega_3 &\leq R_4+R_6 & (\mathbf h_{12})\\2\omega_2+\omega_3 &\leq R_4+ R_5& (\mathbf h_{14})\\2\omega_2+\omega_3 &\leq R_5+R_6 & (\mathbf h_{17})\\ \end{aligned}$   \\
\hline
\end{tabular}
\vspace{3mm}
\caption{Facet orbits of the second symmetric inner bound $\mathcal B_2^G$ obtained by procedure $\textsf{symDD}(\cdot)$.}\label{tab:symsecond_h}
\end{center}
\end{table}

In many contexts, the known symmetry group of the projected polyhedron may not include all of its symmetries.  The next example shows the effect of increasing knowledge of symmetry on a symmetric double descriptions pair updates, the reduction of the number of LPs to solve, and the resulting reduction in run-time of an implementation, when projecting a hypercube.

\begin{figure}[h]
\begin{center}
\includegraphics[width=3in]{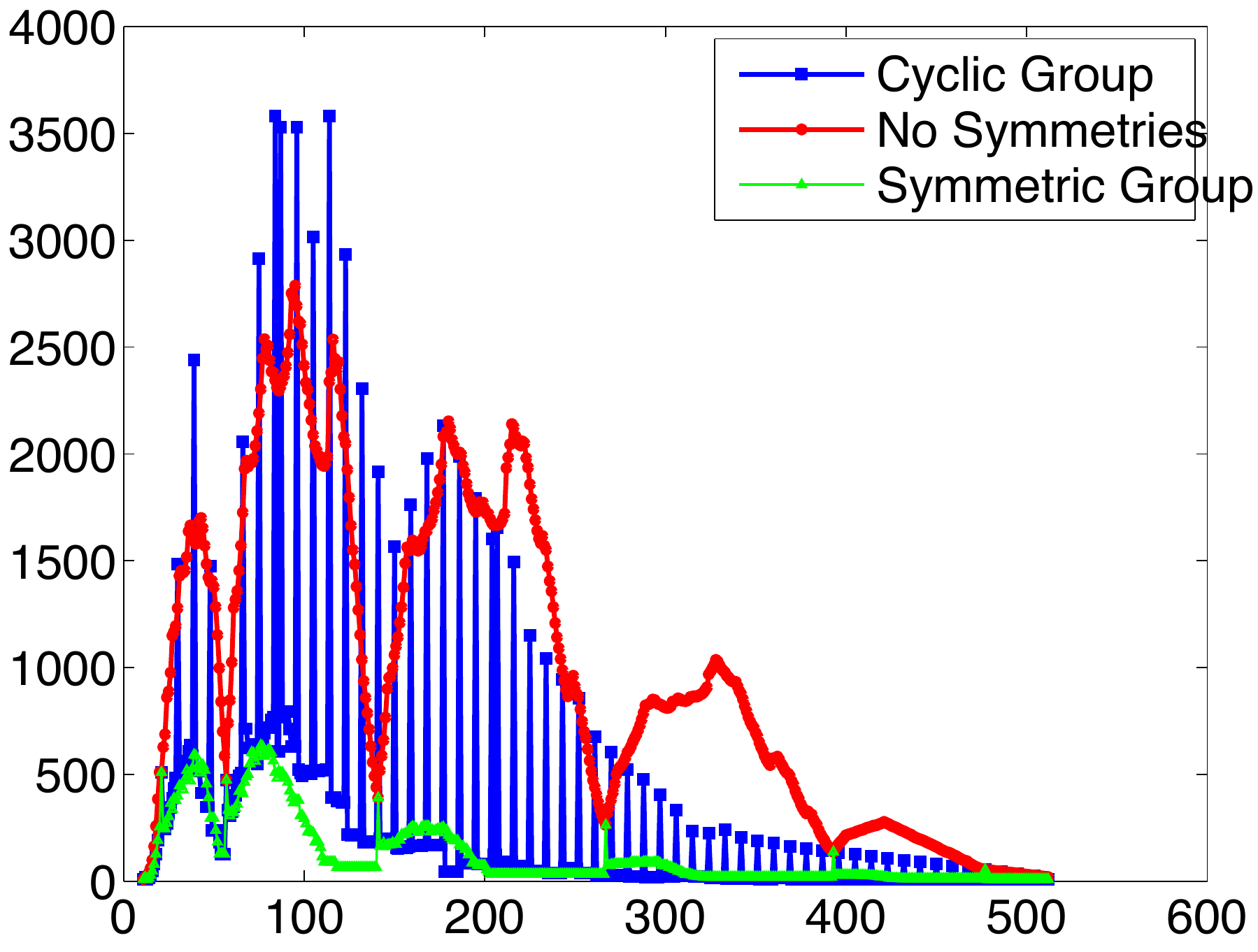}%
\end{center}
\caption{Sizes of double description steps for finding $i$th vertex of the projection of a 12-dimensional hypercube to 9 dimensions versus $i$, under varying knowledge of symmetry.}\label{fig:stepsizeplot}
\end{figure}

\begin{figure}[h]
\begin{center}
\includegraphics[width=3in]{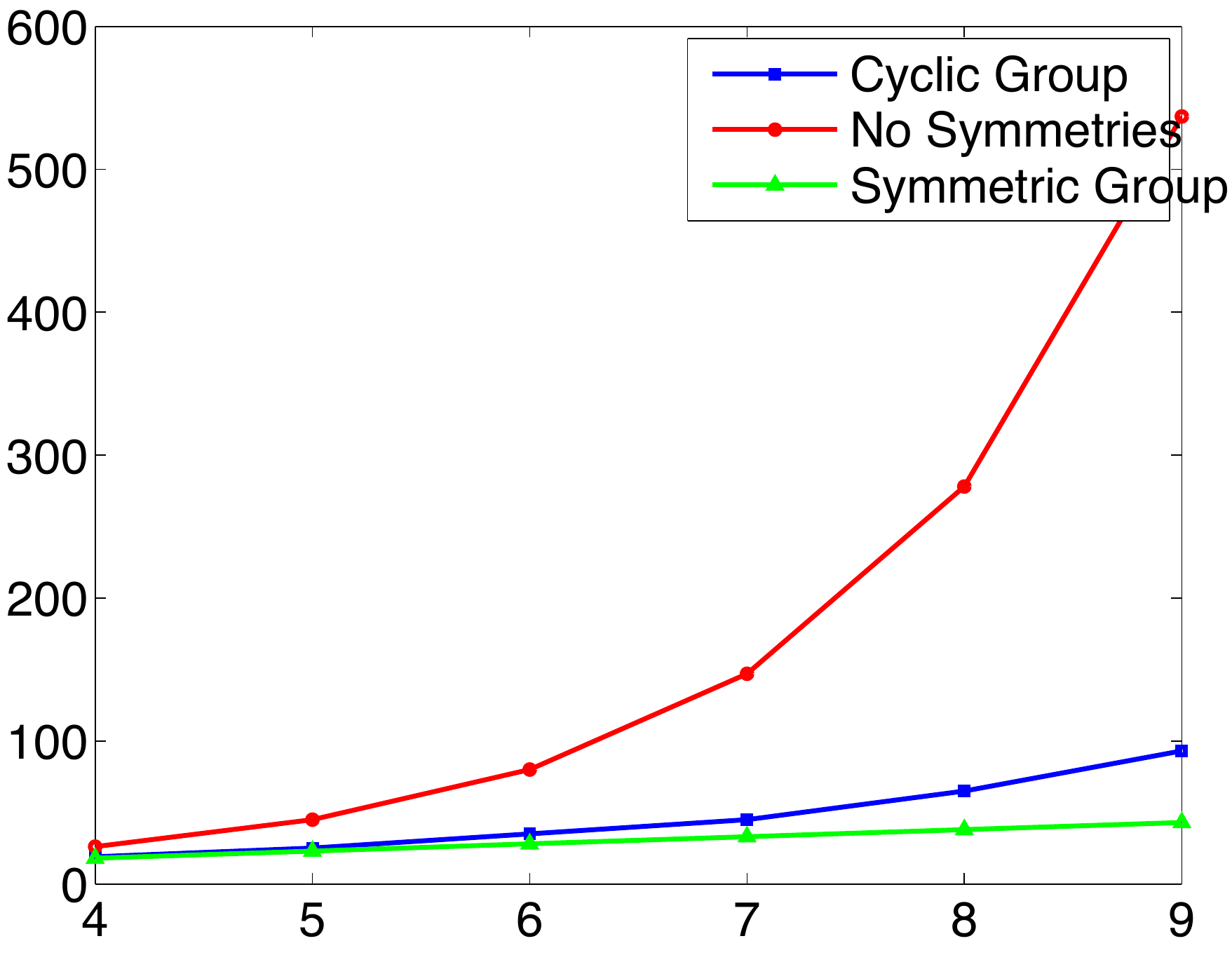}%
\end{center}
\caption{Number of LPs solved for projection of a 12-dimensional hypercube versus the dimension of projection, under varying knowledge of symmetry.}\label{fig:runtimesplot}
\end{figure}

\begin{figure}[h]
\begin{center}
\includegraphics[width=3in]{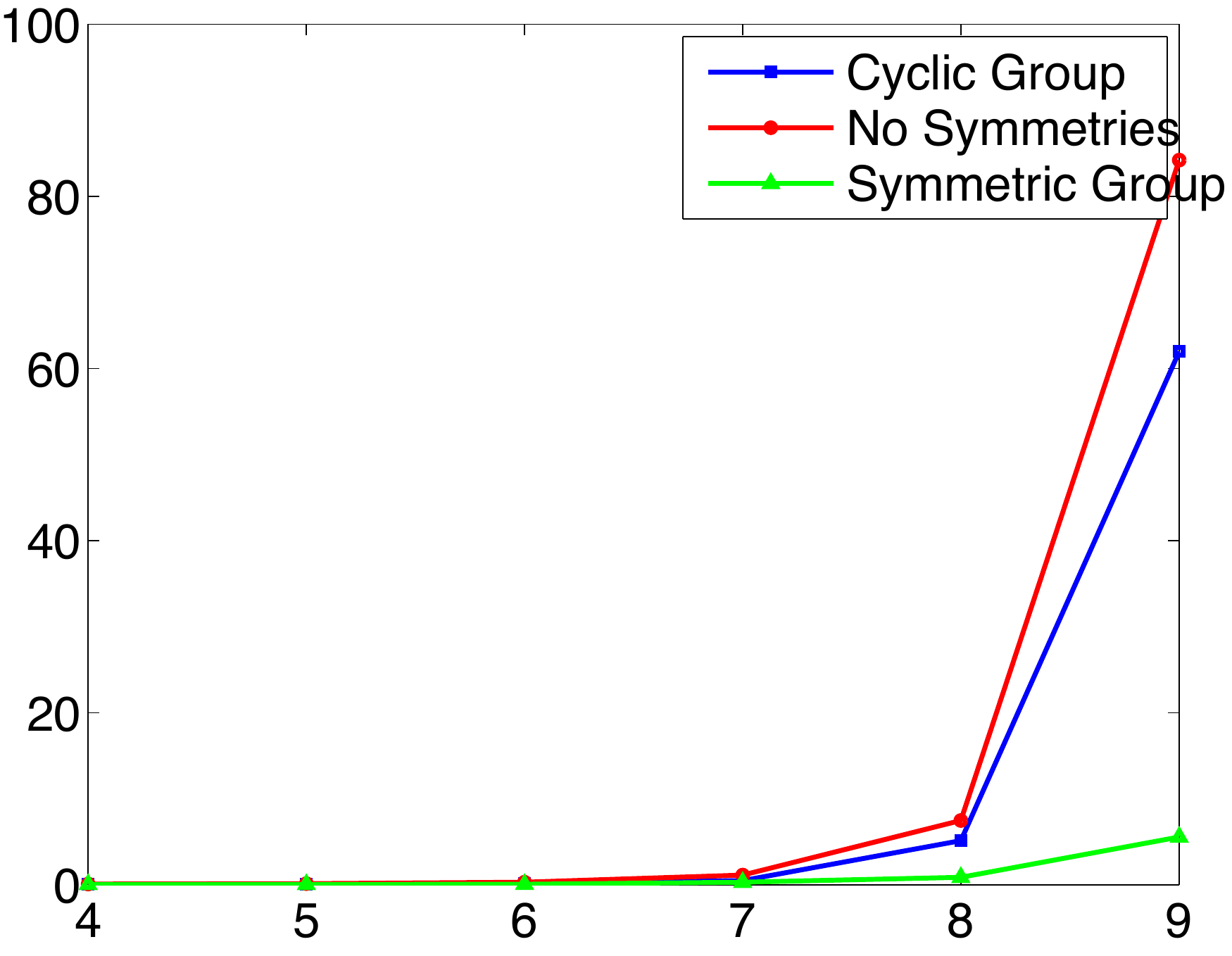}%
\end{center}
\caption{Time is seconds required for projection of a 12-dimensional hypercube versus the dimension of projection, under varying knowledge of symmetry.}\label{fig:numlpplot}
\end{figure}

\exmp{[Varying Symmetry Knowledge when Projecting a Hypercube]
When polytope $\mathcal P$ is a hypercube in $\mathbb R^{12}$, which we want to project down to $\mathbb R^9$, the sizes of DD steps under varying knowledge of symmetry is shown in Fig. \ref{fig:stepsizeplot}. In this case, the number of double description steps does not reduce
i.e. $\vert \mathcal A\vert=\vert\mathbf v^G\vert -1$, in line \ref{ln:repdd} of proc. \textsf{symdd}, for every symmetric update. The main tool to gauge the efficiency of the symmetric updates deescribed in this section, we consider the sizes of extreme ray descriptions input to the double description steps. The average stepsizes under knowledge of no symmetries, cyclic group $C_9$ and symmetric group $S_9$ are $978.97$,  $245.38$ and $120.54$ respectively. Fig. \ref{fig:numlpplot} shows the number of LPs solved under varying knowledge of symmetry for projecting $\mathcal P$ to different dimensions, while Fig. \ref{fig:runtimesplot} shows the time required for projection vs the dimension under varying knowledge of symmetry.}

\subsection{Exploiting Symmetry to Reduce the Dimension of the Linear Program}\label{sec:symLp}
The complexity reductions in the previous two sections are based on exploiting knowledge of the symmetry group of the projected polyhedron that can be inferred from those of the parent polyhedron to project.  However, building on standard symmetry exploitation techniques for symmetric linear programs \cite{herr09arxiv,bodyhighsymmlpalgo}, knowledge of symmetries of the parent polyhedron can also be heavily exploited to reduce the dimension of the linear programs that must be solved.

Taking the subgroup of any known symmetries of the parent polyhedron that fixes the cost in the linear program to solve, one may reduce the dimension of the constraint space (parent polyhedron) by adding the constraints restricting it to the subspace that is fixed under this subgroup without affecting the minimal cost obtained by solving the LP.  The can be executed for both known coordinate permutation symmetries, as well as known affine and restricted affine symmetries.

In the case of symCHM, this capability to reduce the dimension of the LP without affecting the cost can be exploited when checking if a given facet is terminal by solving the associated lower dimensional linear program.  Only in the event that a facet is found to be non-terminal is it necessary to solve a higher dimensional linear program to ensure a new extreme point of the projection's inner bound is revealed.

\exmp{[LP Dimension Reduction for Fig. \ref{fig:sixIDSC}]  By applying these dimensionality reduction using the coordinate permutation symmetries (network symmetry group), one can reduce the dimension of the LP checking for terminal facets when symCHM is applied to the networks in Fig. \ref{fig:sixIDSC} according to the legend in Fig. \ref{fig:symCHMcr}.}

\subsection{Symmetry Exploitation in Weighted Sum-Rate Computation and Related Problems}\label{sec:symLpOth}
An especially attractive application of the dimension reduction trick mentioned in the previous section applies not only to symCHM for full polyhedral projection, but also to simpler problems in network information theory such as weighted sum-rate problems.

Here, we are given an outer bound $\Gamma_{\textsf{out}}$ on the entropy region, a HMSNC instance $\textsf{A}$, weights $\lambda_1,\hdots,\lambda_k$ for each of the $k$ sources and capacities
$r_{k+1},\hdots,r_N$ of the edges. The weighted sum-rate bound associated with $\Gamma_{\textsf{out}}$, is the solution to the following linear program:
\begin{equation}\label{eq:sumratelp}
\max_{\mathcal P} \sum_{i\in [k]}\lambda_i\omega_i
\end{equation}
where,
\begin{equation}
\mathcal P\triangleq \{(\widetilde{\boldsymbol\omega},\widetilde{\mathbf r},\widetilde{\mathbf h})\in  \Gamma_{\textsf{out}}\cap\mathcal L'\cap\mathcal L''| \widetilde{\mathbf r_i}=\mathbf r_i,\forall i\in[N]\setminus [k]\}
\end{equation}
A key consideration that enables the dimension reduction is the symmetries of the following vector,
\begin{equation}\label{eq:lambvec}
\boldsymbol\delta=(\lambda_1,\hdots,\lambda_k,r_{k+1},\hdots,r_N)^T
\end{equation}
under the action of the symmetry group $G$ of  $\Gamma_{\textsf{out}}\cap\mathcal L'\cap\mathcal L''\subseteq \mathbb R^M$. If $G$ is an ASG of $\mathcal P$, let $G'$be the subgroup of $G$ s.t.
\begin{equation}
G'=\left\{ [\mathbf{b},\mathbf{A}] \in G \left|\begin{aligned} &  [\boldsymbol\delta^T \mathbf{0}_{M-N}^T] (\mathbf{A} - \mathbb{I}_{M}) = \mathbf{0}_M,\\
& [\boldsymbol{\delta}^T \mathbf{0}_{M-N}^T] \mathbf{b} = 0 \end{aligned}\right. \right\}
\end{equation}

If $G'$ is the subgroup of $G$ that fixes vector \eqref{eq:lambvec}, the solution of the linear program in equation \eqref{eq:sumratelp} can be obtained by solving the following linear program instead,
\begin{equation}
	\max_{\textsf{Fix}_{G'}(\mathcal P)} \sum_{i\in [k]}\lambda_i\omega_i
\end{equation}
where $\textsf{Fix}_{G'}(\mathcal P)$ is defined as,
\begin{equation}
\textsf{Fix}_{G'}(\mathcal P) = \left\{ \mathbf{y} \in \mathcal P \left| \mathbf{A}\mathbf{y} + \mathbf{b} = \mathbf{y},\  \forall [\mathbf{b},\mathbf{A}] \in G' \right. \right\}.
\end{equation}
The dimension of $\textsf{Fix}_{G'}(\mathcal P)$ is expected to be much smaller than that of $\mathcal P$. For example, if $G'$ is made up of permutation matrices, i.e. it is a subgroup of $S_M$, the dimension of $\textsf{Fix}_{G'}(\mathcal P)$ is equal to number of orbits of $G'$ in set $[M]$.
On the same lines as NSG, one can compute symmetry groups of access structures in secret sharing\cite{padronotes}, and directed graphs in the context of guessing games \cite{riis06}. Computation of lower bounds on worst case information ratio in secret sharing and upper bounds on guessing numbers of graphs, using information inequalities, has the same semantics as the computation of weighted sum-rate bounds in network coding, allowing symmetry to be exploited in identical manner.

\section{The Information Theoretic Converse Prover}  \label{sec:ITCP}
Provided along with this manuscript is a GAP \cite{GAP4} package, the Information Theoretic Converse Prover (ITCP)\cite{jayantitcp}, that implements the described techniques and algorithms.  To ensure that the proofs are exact and not riddled with floating point precision issues, the inbuilt rational arithmetic of GAP from GMP\cite{Granlund12} is exploited along with  exact rational linear programming are achieved via an interface to QSopt\_ex linear programming solver \cite{qsoptex}.   The default outer bound on entropy region, with respect to which EPOBs and other bounds can be computed, is the Shannon outer bound $\Gamma_N$. ITCP also contains an inbuilt database of $215$ non-Shannon inequalities that are inequivalent under random variable permutations, which includes the inequality of Zhang and Yeung \cite{zyineq} and those of Dougherty,Freiling and Zeger \cite{DBLP:journals/corr/abs-1104-3602}.   The user can optionally specify a subset of these inequalities to be considered in the computation.

While the series of examples throughout the manuscript have served the purpose of illustrating the ideas, in this section we aim to demonstrate how to use the GAP package to perform the types of calculations described in the manuscript.  To this end, a series of examples are accompanied with the associated transcripts from sessions with the GAP package.

The first example demonstrates how to compute a network symmetry group with ITCP.
\exmpl{[Computing a Network Symmetry Group with ITCP]  Fig. \ref{fig:itcp1} contains the transcript of a GAP session in which ITCP is utilized to calculate a network symmetry group.  The network coding instance used here is an IDSC \cite{li2015computer} system, that is constructed so that it has a desired symmetry group.  As mentioned in \S\ref{sec:netCodSym}, the network symmetry group is the direct product of two groups the form $G_1\times G_2$ where group $G_1$ is a group of permutations of source labels while group $G_2$ is a group of permutations of edge labels. The instance considered here has size $8$ while its NSG $G$ has order $20$, and is isomorphic to $S_2\times D_5$, where $D5$ is the dihedral group corresponding to the symmetries of a regular pentagon. Starting from group $G$ and $5$ encoders, this instance can be constructed by, 1) adding decoders demanding sources $\{1,2\}$, having access to a set in the orbit of $\{4,5\}$ under $G$, and 2)  adding decoders demanding sources $\{3\}$, having access to a set in the orbit of $\{4,6\}$ under $G$.}{itcpex1}

\begin{figure*}
\begin{Verbatim}[commandchars=!@|,fontsize=\small,frame=single,label=Sample ITCP session for example \ref{itcpex1}]
  !gapprompt@gap>| !gapinput@# Define a size 8 IDSC instance|
  !gapprompt@>| !gapinput@idsc:=[ [ [ [ 1, 2, 3 ], [ 1, 2, 3, 4, 5, 6, 7, 8 ] ],\|
  !gapprompt@>| !gapinput@      [ [ 4, 5 ], [ 1, 2, 4, 5 ] ], [ [ 5, 6 ], [ 1, 2, 5, 6 ] ],\|
  !gapprompt@>| !gapinput@      [ [ 6, 7 ], [ 1, 2, 6, 7 ] ], [ [ 7, 8 ], [ 1, 2, 7, 8 ] ],\|
  !gapprompt@>| !gapinput@      [ [ 4, 8 ], [ 1, 2, 4, 8 ] ], [ [ 4, 6 ], [ 3, 4, 6 ] ],\|
  !gapprompt@>| !gapinput@      [ [ 5, 8 ], [ 3, 5, 8 ] ], [ [ 4, 7 ], [ 3, 4, 7 ] ],\|
  !gapprompt@>| !gapinput@      [ [ 5, 7 ], [ 3, 5, 7 ] ], [ [ 6, 8 ], [ 3, 6, 8 ] ] ], 3, 8 ];|
  !gapprompt@gap>| !gapinput@G:=NetSymGroup(idsc);|
  Group([ (5,8)(6,7), (4,5)(6,8), (4,6)(7,8), (1,2) ])
  !gapprompt@gap>| !gapinput@Size(G);|
  20
\end{Verbatim}
\caption{Example \ref{itcpex1}: Computing a network symmetry group with ITCP.}\label{fig:itcp1}
\end{figure*}

The next example demonstrates how to calculate an information theoretic converse EPOB via ITCP.

\exmpl{[Shannon Outer Bound for the Fano Network]  Fig. \ref{fig:itcpex2} contains the transcript of a GAP session in which ITCP is utilized to calculate the Shannon outer bound to a network coding rate region.  The HMSNC instance considered in this example is the Fano network, which is a well-known matroidal network \cite{DFZMatroidNetworks}, whose network symmetry group is trivial. The EPC is computed using the Shannon outer bound $\Gamma_7$. If we substitute $R_i=1,\forall i\in\{1,\ldots,7\}\setminus \{1,2,3\}$ in the rate region shown in the sample session, we get the region described by Dougherty, Freiling and Zeger in \cite{dougherty2015achievable}, which is a 3 dimensional cube, while the more general list of inequalities describing the region are calculated and given here.}{itcpex2}

\begin{figure*}
\begin{Verbatim}[commandchars=!@|,fontsize=\small,frame=single,label=Sample ITCP session for example \ref{itcpex2}]
  !gapprompt@gap>| !gapinput@# define a network instance (in this case, Fano network)|
  !gapprompt@>| !gapinput@F:=[ [ [ [ 1, 2 ], [ 1, 2, 4 ] ], [ [ 2, 3 ], [ 2, 3, 5 ] ],\|
  !gapprompt@>| !gapinput@      [ [ 4, 5 ], [ 4, 5, 6 ] ], [ [ 3, 4 ], [ 3, 4, 7 ] ],\|
  !gapprompt@>| !gapinput@      [ [ 1, 6 ], [ 3, 1, 6 ] ], [ [ 6, 7 ], [ 2, 6, 7 ] ],\|
  !gapprompt@>| !gapinput@      [ [ 5, 7 ], [ 1, 5, 7 ] ] ], 3, 7 ];;|
  !gapprompt@gap>| !gapinput@rlist:=NCRateRegionOB(F,true,[]);;|
  !gapprompt@gap>| !gapinput@Display(rlist[2]);|
  0 >= -w2
  0 >= -w1
  0 >= -w3
  +R6 >= +w3
  +R5 >= +w3
  +R7 >= +w1
  +R4 >= +w1
  +R6 +R7 >= +w2 +w3
  +R4 +R6 >= +w2 +w3
  +R4 +R5 >= +w2 +w3
  +R6 +R7 >= +w1 +w2
  +R4 +R6 >= +w1 +w2
  +R4 +R5 >= +w1 +w2
\end{Verbatim}
\caption{Example \ref{itcpex2}: Computing the Shannon outer bound on the Fano network, which has the trivial network symmetry group, with polyhedral projection via ITCP.}\label{fig:itcpex2}
\end{figure*}

We next pass to demonstrating how to use ITCP to calculate the rate region for the running example in the manuscript.
\exmpl{[Shannon Outer Bound for Ex. \ref{ex:run}]  Fig. \ref{fig:itcpex3} gives the transcript of a session in GAP in which ITCP is utilized to calculate the Shannon outer bound to a rate region for a symmetric network.  In such cases when NSG is non-trivial, ITCP outputs only one inequality per equivalence class under the action of NSG $G$. There are $18$ such equivalence classes for the IDSC instance under consideration, as shown in the sample ITCP session, while there are a total of 94 distinct permuted forms of these inequalities that form the facets of $\mathcal R_{\textsf{out}}$. A total of $119$ LPs are solved during this computation, while if on decides to ignore symmetries, $207$ LPs must be solved instead. The time required for computation of the EPOBs is 43.91 sec and 67.72 sec, with and without the knowledge of symmetries, respectively.}{itcpex3}

\begin{figure*}
 \begin{Verbatim}[commandchars=!@|,fontsize=\small,frame=single,label=Sample ITCP session for example \ref{itcpex3}]
  !gapprompt@gap>| !gapinput@rlist1:=NCRateRegionOB2(idsc,true,[]);;|
  !gapprompt@gap>| !gapinput@Display(rlist1[2]);|
  0 >= -w2
  0 >= -w3
  +R4 >= 0
  +R4 +R6 >= +w3
  +R4 +R5 >= +w1 +w2
  +R4 +1/2 R5 +1/2 R8 >= +w1 +w2 +1/2 w3
  +1/2 R4 +1/2 R5 +1/2 R6 +1/2 R7 >= +w1 +w2 +1/2 w3
  +2/3 R4 +2/3 R5 +1/3 R6 +1/3 R8 >= +w1 +w2 +2/3 w3
  +2/3 R4 +1/3 R5 +1/3 R6 +1/3 R7 +1/3 R8 >= +w1 +w2 +2/3 w3
  +1/2 R4 +1/2 R5 +1/2 R6 +1/4 R7 +1/4 R8 >= +w1 +w2 +3/4 w3
  +R4 +1/2 R5 +1/2 R6 +1/2 R7 >= +w1 +w2 +w3
  +R4 +1/2 R5 +1/2 R6 +1/2 R8 >= +w1 +w2 +w3
  +R4 +1/3 R5 +1/3 R6 +1/3 R7 +1/3 R8 >= +w1 +w2 +w3
  +2/3 R4 +2/3 R5 +1/3 R6 +2/3 R7 +1/3 R8 >= +w1 +w2 +4/3 w3
  +R4 +1/2 R5 +1/2 R6 +R7 >= +w1 +w2 +3/2 w3
  +R4 +1/2 R5 +1/2 R6 +1/2 R7 +1/2 R8 >= +w1 +w2 +3/2 w3
  +2 R4 +R6 +R7 >= +w1 +w2 +2 w3
  +R4 +R5 +R6 +R7 >= +w1 +w2 +2 w3
\end{Verbatim}
\caption{Example \ref{itcpex3}: Using ITCP to calculate the rate region in Ex. \ref{ex:run}.}\label{fig:itcpex3}
\end{figure*}

In addition to computing explicit polyhedral outer bounds for network coding rate regions, ITCP contains driver routines to calculate upper bounds on the sum rate of a network, lower bounds on the worst case information ratio for a secret sharing problem, and upper bounds on the guessing number of a graph, all of which can exploit problem symmetry in the manner described in \S \ref{sec:symLpOth}.  The remaining three examples demonstrate these capabilities, with the GAP transcript indicating the substantially lower dimensional linear program that is obtained after using the techniques from \S \ref{sec:symLpOth}.

\exmpl{[Upper Bound on a Network Coding Sum Rate with ITCP]  Fig. \ref{fig:itcpex4} gives the transcript of a GAP session in which ITCP is utilized to upper bound the sum rate for a network coding problem.   The selected problem is a size $5$ HMSNC instance, with symmetry group of order $2$, that contains permutations $\{(1),(3,4)\}$. By exploiting these symmetries, the dimension of the linear program can be reduced from 28 to 22.}{itcpex4}

\begin{figure*}
\begin{Verbatim}[commandchars=!@|,fontsize=\small,frame=single,label=Sample ITCP session for example \ref{itcpex4}]
  !gapprompt@gap>| !gapinput@# define a network instance|
  !gapprompt@>| !gapinput@N:= [[ [ [ 1 ], [ 1, 3 ] ], [ [ 1 ], [ 1, 4 ] ],[ [ 1, 2, 5 ],\|
  !gapprompt@>| !gapinput@      [ 1, 2 ] ],[ [ 1, 2, 3 ], [ 2, 3 ] ],[ [ 2, 4 ], [ 1, 2, 4 ] ],\|
  !gapprompt@>| !gapinput@      [ [ 2, 3, 4, 5 ], [ 3, 4, 5 ] ]] , 2, 5 ];;|
  !gapprompt@gap>| !gapinput@ub:=NCSumRateUB(N,[1,1,1],[]);;|
  Original LP dimension...28
  LP dimension after considering symmetries...22
  !gapprompt@gap>| !gapinput@ub;|
  2
\end{Verbatim}
\caption{Example \ref{itcpex4}: computing a network coding sum rate bound using ITCP.}\label{fig:itcpex4}
\end{figure*}

\exmpl{[Secret Sharing Information Ratio Bound via ITCP] Fig. \ref{fig:itcpex5} gives a transcript of a GAP session in which ITCP is used to lower bound the worst case information ratio lower bounds for a specified secret sharing access structure. The access structure considered here has authorized sets $\{\{2,3\},\{3,4\},\{4,5\}\}$, where there are 4 parties labeled $\{2,3,4,5\}$ while the dealer of secret is labeled $1$. The lower bound is computed with respect to $\Gamma_N$, which is already known in the literature to be $\frac{3}{2}$, which  is also achievable using multi-linear scheme (see \cite{padronotes} \S 2.8).  The symmetry detected in the problem decreases the dimension of the associated linear program from 20 to 12.}{itcpex5}

\begin{figure*}
\begin{Verbatim}[commandchars=!@|,fontsize=\small,frame=single,label=Sample ITCP session for example \ref{itcpex5}]
  !gapprompt@gap>| !gapinput@# define an access structure|
  !gapprompt@>| !gapinput@Asets:=[[2,3],[3,4],[4,5]];;|
  !gapprompt@gap>| !gapinput@lb:=SSWorstInfoRatioLB(Asets,5,[]);;|
  Original LP dimension...20
  LP dimension after considering symmetries...12
  !gapprompt@gap>| !gapinput@lb;|
  3/2
\end{Verbatim}
\caption{Example \ref{itcpex5}: computing a secret sharing information ratio bound with ITCP.}\label{fig:itcpex5}
\end{figure*}

\exmpl{[Bounding the Guessing Number of a Graph with ITCP] Fig. \ref{fig:itcpex5} gives a GAP transcript of a session wherein ITCP is used to determine upper bounds on the guessing number of a directed graph. The graph under consideration is the cycle graph $C_5$. The upper bound obtained using $\Gamma_5$, in this case, is $\frac{5}{2}$ (see eg. \cite{atkinsguessoddcycle}).  The problem dimension is reduced from 25 to 5 by exploiting symmetry.}{itcpex6}

\begin{figure*}
\begin{Verbatim}[commandchars=!@|,fontsize=\small,frame=single,label=Sample ITCP session for example \ref{itcpex6}]
  !gapprompt@gap>| !gapinput@# define a directed graph (in this case, the cycle graph C5)|
  !gapprompt@>| !gapinput@C5:=[ [ 1, 2, 3, 4, 5 ],\|
  !gapprompt@>| !gapinput@  rec( 1 := [ 2, 5 ], 2 := [ 1, 3 ], 3 := [ 2, 4 ],\|
  !gapprompt@>| !gapinput@      4 := [ 3, 5 ], 5 := [ 4, 1 ] ) ];;|
  !gapprompt@gap>| !gapinput@ub:=GGnumberUB(C5,[]);;|
  Original LP dimension...25
  LP dimension after considering symmetries...5
  !gapprompt@gap>| !gapinput@ub;|
  5/2
\end{Verbatim}
\caption{Example \ref{itcpex6}: computing an upper bound on the guessing number of a directed graph.}\label{fig:itcpex6}
\end{figure*}

\section{Conclusions}

\bibliographystyle{IEEEtran}
\bibliography{IEEEabrv,NETCOD2015,myPubsWLinks}

\end{document}